\documentclass[aps,prb,reprint,superscriptaddress]{revtex4-1}
\usepackage{graphicx}
\usepackage{dcolumn}
\usepackage{bm}
\usepackage{hyperref}
\usepackage{amssymb}
\usepackage{amsmath}
\usepackage{epsf}
\usepackage[caption=false]{subfig}
\usepackage{epstopdf}
\usepackage{amsfonts}
\usepackage{color}

\usepackage{pst-all}
\begin{document}
\title{Counting statistics of energy transport across squeezed thermal 
reservoirs}
\author{Hari Kumar Yadalam}
\email{hari.kumar@icts.res.in}
\affiliation{International Centre for Theoretical Sciences, Tata Institute of 
Fundamental Research, 560089 Bangalore, India}
\affiliation{Laboratoire de Physique, \'Ecole Normale Sup\'erieure, CNRS, 
Universit\'e
PSL, Sorbonne Universit\'e, Universit\'e de Paris, 75005 Paris, France}
\author{Bijay Kumar Agarwalla}
\email{bijay@iiserpune.ac.in}
\affiliation{Department of Physics, Indian Institute of Science Education and 
Research Pune, Dr. Homi Bhabha Road, Ward No. 8, NCL Colony, Pashan, Pune, 
Maharashtra 411008, India}
\author{Upendra Harbola}
\email{uharbola@iisc.ac.in}
\affiliation{Department of Inorganic and Physical Chemistry, Indian Institute 
of Science, Bangalore, 560012, India}
\date{\today}
\keywords{Fluctuation Theorems, Full Counting Statistics, Quantum Master 
Equations, Phase Space Quasiprobability Distribution Functions, Quantum 
Optics, Squeezed States, Nonequilibrium Statistical Mechanics} 
\begin{abstract}
A general formalism for computing the full counting statistics 
of energy exchanged between 'N' squeezed thermal photon reservoirs weakly 
coupled to a cavity  with 'M' photon modes is presented. The formalism is 
based on the two-point measurement scheme and is applied to two simple special 
cases, the relaxation dynamics of a single mode 
cavity in contact with a single squeezed thermal photon reservoir and the 
steady-state energy transport between two squeezed thermal photon reservoirs 
coupled to a single cavity mode. Using analytical results, it is found that the 
short time statistics is significantly affected by  noncommutivity of the 
initial energy measurements with the reservoirs squeezed states, and may 
lead to negative probabilities if not accounted properly. 
Furthermore, it is found that for the single reservoir setup, generically there 
is no transient or steady-state fluctuation theorems for energy transport. In 
contrast, for the two reservoir case, although there is no generic transient 
fluctuation theorem, steady-state fluctuation theorem with a non-universal 
affinity is found to be valid. Statistics of energy currents are further 
discussed.
\end{abstract}
\maketitle
\section{Introduction}
Fluctuations of observables in physical systems are ubiquitous. These 
fluctuations, seemingly arbitrary, carry a great amount of information related 
to the underlying physical processes. For example fluctuations of 
observables in systems at equilibrium are known to be related to 
their responses to weak perturbations through fluctuation-dissipation theorem. 
These relations are valid only for systems close to equilibrium 
\cite{callen1951irreversibility,green1954markoff,kubo1957statistical,de2013non}. 
Towards the end of the twentieth 
century, the past three decades of research, fluctuations in physical 
systems, even far from equilibrium, under certain conditions, were shown to 
satisfy universal relations, dubbed as fluctuation theorems 
\cite{Esposito2009,Campisi2011,Seifert2012,Klages2013}. These 
fluctuation theorems, have been demonstrated 
for various non-equilibrium systems, such as heat and charge transport in 
nano-meter sized junctions like nanoelectronic quantum dot junctions, molecular 
junctions, cavity photonic systems, nano sized hybrid electro-optical, and 
electromechanical systems. 

The fluctuation theorems are microscopic expressions 
of second law of thermodynamics and are derived based on the 
assumption that system's initial state is canonical (local) equilibrium state 
and the dynamics is micro-reversible 
\cite{Esposito2009,Campisi2011,Seifert2012,Klages2013}. To our knowledge not 
much work has been done in exploring the existence of fluctuation 
theorems for specially prepared non-canonical initial states. One such special 
class of non-canonical states of recent interest has been squeezed thermal 
states of photons. Squeezed thermal states of bosonic reservoirs have been 
used to enhance the efficiency of heat engines  \cite{giraldi2014coherence}. It 
was shown that quantum heat engines with squeezed reservoirs can have efficiency 
more than the Carnot efficiency \cite{Huang2012,Abah2014,Rossnagel2014} and 
allow work extraction even from a single reservoir \cite{Manzano2016}. Later 
works have established generalized Carnot 
type bounds on the efficiencies of engines with squeezed reservoirs 
\cite{Alicki2015,Niedenzu2016,Agarwalla2017,Niedenzu2018}. Some of these 
predictions have been realized in a recent experiment \cite{Klaers2017}.

However, it is not clear how any of the established fluctuation theorems 
\cite{Esposito2009,Campisi2011,Seifert2012,Klages2013} are modified for systems 
prepared in non-canonical states and whether there is a form 
of fluctuation theorem, transient or steady-state. Motivated by these 
questions, in this work we study statistics of energy transport and explore 
the question of existence of fluctuation theorem in very simple model system 
consisting of a 'M' photon modes of a cavity coupled to 'N' squeezed thermal 
photon reservoirs. It is important to note that for a qubit system coupled to 
squeezed thermal reservoir, reservoir can be characterized using an effective 
temperature and an effective fluctuation theorem may be valid 
\cite{Agarwalla2017}. However, it is not clear if this is a generic feature or a 
result of special system under consideration. As we discuss in this work, qubit 
system indeed is a possible exception. It is also to be noted that even for 
canonical reservoirs with micro-reversible dynamics, new fluctuation theorems, 
different from traditional ones, can emerge for particle currents through 
superconducting systems as a result of the $\mathcal{U}(1)$ symmetry breaking, 
particle number non-conserving terms, in the microscopic Hamiltonians 
\cite{zhang2021full}.

This work is organized as follows. After introducing the model system in 
Sec. (\ref{sec2}), the description and computation of the moment generating 
function are presented in Sec. (\ref{sec3}). These are then followed by the 
application of the results to two simple model systems in Sec. (\ref{sec4}). 
Finally conclusions are presented. Few details of the computations are 
relegated to the appendix.   

\section{Model system}
\label{sec2}
\begin{figure}[tbh!]
\centering
\includegraphics[width=6.5cm,height=4.2cm]{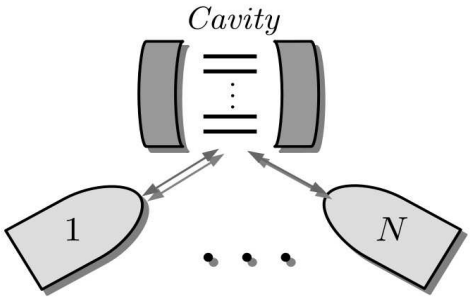}
\caption{Schematic of the model considered. The model consists of a '$M$' mode 
cavity prepared in squeezed thermal state coupled to '$N$' reservoirs prepared 
in squeezed thermal states.}
\label{model}
\end{figure}

The model system considered in this work consists of a cavity having $M$ photon 
modes, weakly coupled to $N$ photon reservoirs. The Hamiltonian describing the 
system is, 
\begin{eqnarray}
\label{eq-1}
 H_{}^{}&=& \underbrace{\sum_{i,j = 1}^{M} b_{S i}^{\dag} {h_{S}^{}}_{ij}^{} 
b_{S j}^{}}_{H_{S}^{}} + \sum_{\alpha=1}^{N}\underbrace{\sum_{k \in 
\alpha}^{}\epsilon_{\alpha k}^{} b_{\alpha k}^\dag b_{\alpha 
k}^{}}_{H_{\alpha}^{}}\nonumber\\ 
 &&+ i\sum_{\alpha=1}^{N}\underbrace{\sum_{k \in \alpha}^{}\sum_{i=1}^{M} g_{S i 
\alpha k}^{} \left[b_{\alpha k}^\dag b_{S i}^{} - b_{S i}^\dag b_{\alpha 
k}^{}\right]}_{H_{S \alpha}^{}}.
\end{eqnarray}
Here  $b_{S i}^{\dag}$ ($b_{S i}^{}$) and $b_{\alpha k}^\dag$ ($b_{\alpha 
k}^{}$) are bosonic creation (annihilation) operators for creating 
(annihilating) a 
photon in the '$i^{th}_{}$' cavity mode and in the '$k_{}^{\text{th}}$' mode in 
the $\alpha^{th}_{}$ photonic reservoir, respectively, and ${h_{S}^{}}_{ij}^{} 
= \epsilon_{S i}^{} \delta_{ij}^{}$. 
Schematic of the model considered is displayed in Fig. (\ref{model}). 

Initially, at time $t=0$, it is assumed that the cavity photon modes and the 
photon reservoirs are not coupled  and are prepared in individual squeezed 
thermal states, i.e., the full density matrix of the whole system at initial 
time is assumed to be of the product (uncorrelated) form,  
\begin{eqnarray}
\label{eq-2}
 \rho_{}^{}(0) &=& \rho_{S}^{}(0)\otimes_{\alpha = 1}^{N}\rho_{\alpha}^{}(0),
\end{eqnarray}
where
\begin{eqnarray}
\label{eq-3}
\rho_{\alpha}^{}(0)=S_{\alpha}^{\dag}\frac{e^{-\beta_{\alpha}^{}H_{\alpha}^{}}_{
} } { \mathbf { Tr } [ e^ { -\beta_{\alpha}^{}
H_{\alpha}^{}}_{}]}S_{\alpha}^{},
\end{eqnarray}
for $\alpha=S, 1, \cdots, N$ and 
\begin{eqnarray}
 S_{\alpha}^{} &=& e^{-\frac{1}{2}\underset{r 
\in \alpha}{\sum}_{}^{}Z_{\alpha}^{}[e_{}^{i \phi_{\alpha}^{}}{b_{\alpha 
r}^{\dag 2}}-e_{}^{-i \phi_{\alpha}^{}}{b_{\alpha 
r}^{2}}]}_{},\nonumber
\end{eqnarray}
being the squeezing operator 
\cite{Scully1997,Agarwal2013,Garrison2014,Lvovsky2015}. For the sake of 
simplicity it is assumed that the squeezing amplitude $Z_{\alpha}^{} \geq 0$ 
and the phase $\phi_{\alpha}^{} \in [-\pi,+\pi)$ of each subsystem (i.e., 
system and reservoirs) are mode-independent. 

In order to study fluctuations of energy transfer from the system into squeezed 
thermal reservoirs, in the next section, we construct full distribution of 
energy transfer using two-point measurement scheme 
\cite{Esposito2009,Campisi2011,Kurchan2000,Tasaki2000,Monnai2005} for the 
system depicted in Fig. (\ref{model}). 
\section{Moment generating function}
\label{sec3}
The cavity and the reservoirs prepared in uncorrelated squeezed thermal states 
are coupled at time $t=0$ (by turning on $H_{S \alpha}^{}$) leading to the flow 
of energy between the system and the reservoirs. The joint probability 
distribution for the amount of energy flowing, $\mathbf{\Delta e}_{}^{} = 
\begin{pmatrix} \Delta e_{1}^{} & \cdots & \Delta e_{N}^{}\end{pmatrix}_{}^{T}$, 
into each of the reservoirs in time $t_{}^{}$, can be written as,
\begin{eqnarray}
\label{eq-4}
 &&P[\mathbf{\Delta 
e}_{}^{},t_{}^{}]=\frac{1}{(2\pi)_{}^{N}}\underset{\bm{\chi}_{}^{} \in 
\mathbb{R}_{}^{N}}{\int}_{}^{} d_{}^{N}\bm{\chi}_{}^{}\ 
 \mathcal{Z}[\bm{\chi}_{}^{},t_{}^{}]e_{}^{i \bm{\chi}_{}^{T}\mathbf{\Delta 
e}_{}^{}}
\end{eqnarray}
where $\mathcal{Z}[\bm{\chi}_{}^{},t_{}^{}]$ is the moment generating function 
which within the two-point measurement scheme 
\cite{Esposito2009,Campisi2011,Kurchan2000,Tasaki2000,Monnai2005} is obtained 
as,
\begin{eqnarray}
\label{eq-5}
\mathcal{Z}[\bm{\chi}_{}^{},t_{}^{}]=\frac{1}{(2\pi)_{}^{N}}\underset{\bm{
\lambda}_{}^{} \in \mathbb{R}_{}^{N}}{\int}_{}^{} d_{}^{N}\bm{\lambda}_{}^{}\  
 \tilde{\mathcal{Z}}[\bm{\chi}_{}^{},\bm{\lambda}_{}^{},t_{}^{}]
\end{eqnarray}
with
\begin{eqnarray}
\label{eq-6}
&&\tilde{\mathcal{Z}}[\bm{\chi}_{}^{},\bm{\lambda}_{}^{},t_{}^{}]=\mathbf{Tr}_{
S 
+ B}^{}
\left[e^{-\frac{i}{\hbar} 
H[\bm{\lambda}_{}^{}+\frac{1}{2}\bm{\chi}_{}^{}]t_{}^{}}_{}\rho_{}^{}(0)e^{\frac
{i}{\hbar} 
H[\bm{\lambda}_{}^{}-\frac{1}{2}\bm{\chi}_{}^{}]t_{}^{}}_{}\right],
\end{eqnarray}
where $\bm{\chi}_{}^{} =\begin{pmatrix} \chi_{1}^{} & \cdots & 
\chi_{N}^{}\end{pmatrix}_{}^{T}$ keeps track of the energy flow, 
$\mathbf{\Delta e}_{}^{}$, from the system into the reservoirs, and 
$\bm{\lambda}_{}^{} = \begin{pmatrix} \lambda_{1}^{} & \cdots & 
\lambda_{N}^{}\end{pmatrix}_{}^{T}$ 
carries the information of the initial projective measurement of energy of the 
reservoirs. The integral over  $\bm{\lambda}_{}^{}$ in Eq. (\ref{eq-5}) is 
necessary because the initial density matrices of the reservoirs do not commute 
with the initial projective energy measurements on the reservoirs. This 
integral essentially projects out the initial coherences between isolated 
reservoirs energy eigenstates which are destroyed by the initial projective 
measurements on the reservoirs \cite{Agarwalla2012}. It is crucial to note that 
the above procedure of implementing initial projections should be treated with 
caution, as naively using 
$\tilde{\mathcal{Z}}[\bm{0}_{}^{},\bm{\lambda}_{}^{},t_{}^{}]=1$ from Eq. 
(\ref{eq-6}) in Eq. (\ref{eq-5}) leads to divergence. However it can be made 
meaningful by a physical limiting procedure discussed at the end of this 
section. 

The counting-field-dependent Hamiltonian of the whole system in Eq. 
(\ref{eq-6}) is defined as,
\begin{eqnarray}
\label{eq-7}
&&H[\bm{\chi}_{}^{}]=\sum_{i,j = 1}^{M} b_{S i}^{\dag} {h_{S}^{}}_{ij}^{} b_{S 
j}^{} + \sum_{\alpha=1}^{N}\sum_{k \in \alpha}^{}\epsilon_{\alpha k}^{} 
b_{\alpha k}^\dag b_{\alpha k}^{}\nonumber\\ 
&&+ i\sum_{\alpha=1}^{N}\sum_{k \in \alpha}^{}\sum_{i=1}^{M} g_{S i \alpha 
k}^{}\left[e_{}^{-i \epsilon_{\alpha k}^{} \chi_{\alpha}^{}}b_{\alpha k}^\dag 
b_{S 
i}^{} - e_{}^{i \epsilon_{\alpha k}^{} \chi_{\alpha}^{}} b_{S i}^\dag b_{\alpha 
k}^{}\right].
\nonumber\\
\end{eqnarray}

$\tilde{\mathcal{Z}}[\bm{\chi}_{}^{},\bm{\lambda}_{}^{},t_{}^{}]$ defined in 
Eq. 
(\ref{eq-6}) can be recast as,
\begin{eqnarray}
\label{eq-8}
\tilde{\mathcal{Z}}[\bm{\chi}_{}^{},\bm{\lambda}_{}^{},t_{}^{}]=\mathbf{Tr}_{S}^
{}\left[\rho_{S}^{}(t)\right]
\end{eqnarray}
with the counting-field dependent system's reduced density matrix 
($\rho_{S}^{}(t)$) in the interaction picture, defined as,
\begin{equation}
\label{eq-9}
 \rho_{S}^{}(t)=e_{}^{\frac{i}{\hbar} H_{S}^{} t}\mathbf{Tr}_{B}^{}
 \left[e^{-\frac{i}{\hbar} 
H[\bm{\lambda}_{}^{}+\frac{1}{2}\bm{\chi}_{}^{}]t_{}^{}}_{}\rho_{}^{}(0)e^{\frac
{i}{\hbar} 
H[\bm{\lambda}_{}^{}-\frac{1}{2}\bm{\chi}_{}^{}]t_{}^{}}_{}\right]e_{}^{-\frac{i
}{\hbar} H_{S}^{} t}.\nonumber\\
\end{equation}

By invoking Born-Markov-Secular approximations (and also neglecting the Lamb 
shifts), counting-field dependent Lindblad quantum master equation can be 
derived for $\rho_{S}^{}(t)$ 
\cite{Scully1997,Breuer2002,Carmichael2003,Bagrets2003,Harbola2006,Harbola2007,
Carmichael2009}. This is given as,
\begin{widetext}
\begin{eqnarray}
\label{eq-10}
\frac{\partial}{\partial t}\rho_{S}^{}(t)&=&- \sum_{\alpha = 1}^{N} 
\mathbf{B}_{S}^{T} \left\{ e_{}^{i \mathbf{h}_{S}^{} \left(\lambda_{\alpha}^{} 
+ 
\frac{1}{2}\chi_{\alpha}^{}\right)} \mathbf{\Gamma}_{\alpha}^{} 
\bm{\sigma}_{y}^{} \left[\mathbf{D}_{\alpha}^{} + \frac{i}{2} 
\bm{\sigma}_{y}^{}\right]\bm{\sigma}_{y}^{} e_{}^{i \mathbf{h}_{S}^{} 
\left(\lambda_{\alpha}^{} - \frac{1}{2}\chi_{\alpha}^{}\right)} \right\} 
\rho_{S}^{}(t)\mathbf{B}_{S}^{} \nonumber\\
&&+\frac{1}{2}\sum_{\alpha = 1}^{N} \mathbf{B}_{S}^{T}\left\{ e_{}^{i 
\mathbf{h}_{S}^{} \left(\lambda_{\alpha}^{} + 
\frac{1}{2}\chi_{\alpha}^{}\right)} \mathbf{\Gamma}_{\alpha}^{} 
\bm{\sigma}_{y}^{} \left[\mathbf{D}_{\alpha}^{} - \frac{i}{2} 
\bm{\sigma}_{y}^{}\right]\bm{\sigma}_{y}^{} e_{}^{i \mathbf{h}_{S}^{} 
\left(\lambda_{\alpha}^{} + \frac{1}{2}\chi_{\alpha}^{}\right)}\right\} 
\mathbf{B}_{S}^{} \rho_{S}^{}(t) \nonumber\\
&&+\frac{1}{2}\sum_{\alpha = 1}^{N} \rho_{S}^{}(t) \mathbf{B}_{S}^{T} \left\{ 
e_{}^{i \mathbf{h}_{S}^{} \left(\lambda_{\alpha}^{} - 
\frac{1}{2}\chi_{\alpha}^{}\right)} \mathbf{\Gamma}_{\alpha}^{} 
\bm{\sigma}_{y}^{} \left[\mathbf{D}_{\alpha}^{} - \frac{i}{2} 
\bm{\sigma}_{y}^{}\right]\bm{\sigma}_{y}^{} e_{}^{i \mathbf{h}_{S}^{} 
\left(\lambda_{\alpha}^{} - \frac{1}{2}\chi_{\alpha}^{}\right)} \right\} 
\mathbf{B}_{S}^{},
\end{eqnarray}
\end{widetext}
where $\mathbf{B}_{S}^{} = \begin{pmatrix} b_{S 1}^{\dag} & \cdots & b_{S 
M}^{\dag} & b_{S 1}^{} & \cdots & b_{S M}^{} \end{pmatrix}_{}^{T}$, 
$\mathbf{h}_{S}^{} = \sigma_{z}^{} \otimes h_{S}^{}$, 
$\bm{\sigma}_{x,y,z}^{} = \sigma_{x,y,z}^{} \otimes I_{M \times 
M}^{}$, with $\sigma_{x,y,z}^{}$ being Pauli matrices and $I_{M \times 
M}^{}$ being the $M \times M$ identity matrix, 
$\mathbf{\Gamma}_{\alpha}^{} = I_{}^{} \otimes \Gamma_{\alpha}^{}$ (for 
brevity $I_{2 \times 2}^{}$ is denoted by $I_{}^{}$) with
\begin{eqnarray}
\label{eq-11}
&&{\Gamma_{\alpha}^{}}_{ij}^{}= \begin{cases} \frac{2\pi}{\hbar}\sum_{k \in 
\alpha}^{} g_{S i \alpha k} g_{S j \alpha k}\delta(\epsilon_{\alpha 
k}^{}-\epsilon_{S i}^{}) & \text{if}\ \epsilon_{S i}^{} = \epsilon_{S j}^{} \\ 
0 
& \text{if}\ \epsilon_{S i}^{} \neq \epsilon_{S j}^{} \end{cases},\nonumber
\end{eqnarray}
and $\mathbf{D}_{\alpha}^{} =  -i \bm{\sigma}_{y}^{} e_{}^{-i 
\mathbf{S}_{\alpha}^{} \bm{\sigma}_{y}^{}}\left[ 
n_{\alpha}^{}(\mathbf{h}_{S}^{}) + \frac{1}{2} \mathbf{I}_{}^{}\right]e_{}^{i 
\mathbf{S}_{\alpha}^{} \bm{\sigma}_{y}^{}}$ ($\mathbf{I}_{}^{} = I_{2 M \times 
2 
M}^{}$)  
with $\mathbf{S}_{\alpha}^{} = Z_{\alpha}^{} \bm{\sigma}_{z}^{} e_{}^{i 
\bm{\sigma}_{z}^{} \phi_{\alpha}^{}}$, 
$n_{\alpha}^{}(x)=\left(e_{}^{\beta_{\alpha}^{} x}-1\right)_{}^{-1}$.
The solution of Eq. (\ref{eq-10}), supplemented with the initial 
condition $\rho_{S}^{}(t)|_{t = 0}^{} = \rho_{S}^{}(0)$, when used in Eq. 
(\ref{eq-8}) gives 
$\tilde{\mathcal{Z}}[\bm{\chi}_{}^{},\bm{\lambda}_{}^{},t_{}^{}]$.

Instead of solving the above equation for $\rho_{S}^{}(t)$, we find 
it convenient to solve for the counting-field dependent Wigner 
function, $\mathbb{P}_{}^{}[\bm{\Upsilon}_{}^{},t]$ [$\bm{\Upsilon}_{}^{} 
= \begin{pmatrix} \gamma_{1}^{*} & \cdots & \gamma_{M}^{*} & \gamma_{1}^{} & 
\cdots & \gamma_{M}^{} \end{pmatrix}_{}^{T}$], in the interaction picture 
\cite{Lee1995,Hillery1984,Scully1997,Carmichael2003,Zachos2005,Carmichael2009, 
Schleich2011,Curtright2013}. This is defined as the Fourier transform of the 
Weyl (symmetric ordered moment) generating function 
\cite{Lee1995,Hillery1984,Scully1997,Carmichael2003,Zachos2005,Carmichael2009,
Schleich2011} for the system,
\begin{eqnarray}
\label{eq-13}
 \mathbb{P}_{}^{}[\bm{\Upsilon}_{}^{},t]=\frac{1}{\pi_{}^{2 
M}}\int\mathcal{D}[\mathbf{W}_{}^{}] \underbrace{\mathbf{Tr}_{S}^{}\left[e^{i 
\mathbf{W}_{}^{\dag} 
\mathbf{B}_{S}^{}}_{}\rho_{S}^{}(t_{}^{})\right]}_{\text{Weyl generating 
function}} e^{-i \mathbf{W}_{}^{\dag}\bm{\Upsilon}_{}^{}}_{},\nonumber\\
\end{eqnarray}
where $\mathbf{W}_{}^{} = \begin{pmatrix} w_{1}^{*} & \cdots & w_{M}^{*} & 
w_{1}^{} & \cdots & w_{M}^{} \end{pmatrix}_{}^{T}$.  

$\tilde{\mathcal{Z}}[\bm{\chi}_{}^{},\bm{\lambda}_{}^{},t_{}^{}]$ is then 
expressed in terms of $\mathbb{P}_{}^{}[\bm{\Upsilon}_{}^{},t]$ as,
\begin{eqnarray}
\label{eq-12} 
\tilde{\mathcal{Z}}[\bm{\chi}_{}^{},\bm{\lambda}_{}^{},t_{}^{}]=\int_{}\mathcal{
D}[\bm{\Upsilon}_{}^{}]\ \mathbb{P}_{}^{}[\bm{\Upsilon}_{}^{},t],
\end{eqnarray}
where the short hand notation 
$\int_{}\mathcal{D}[\bm{\Upsilon}_{}^{}]=\int_{-\infty}^{+\infty}d[\Re(\gamma_{1
}^{})\int_{-\infty}^{+\infty}d[\Im(\gamma_{1}^{})]\cdots\int_{-\infty}^{+\infty}
d[\Re(\gamma_{M}^{})\int_{-\infty}^{+\infty}d[\Im(\gamma_{M}^{})]$ is 
introduced.

Using the Lindblad quantum master equation given in Eq. (\ref{eq-10}), 
evolution equation for the Wigner function, 
$\mathbb{P}_{}^{}[\bm{\Upsilon}_{}^{},t]$, is obtained as 
\cite{Scully1997,Carmichael2003,Carmichael2009,Curtright2013},
 \begin{eqnarray}
 \label{eq-14}
  \frac{\partial}{\partial 
t}\mathbb{P}_{}^{}[\bm{\Upsilon}_{}^{},t]&=&\frac{1}{2}
  \left[ \begin{pmatrix} \bm{\Upsilon}_{}^{} \\ 
\bm{\nabla}_{\bm{\Upsilon}_{}^{}}^{} \end{pmatrix}^{T}_{}
  \mathbb{H}_{}^{}
  \begin{pmatrix} \bm{\Upsilon}_{}^{} \\ \bm{\nabla}_{\bm{\Upsilon}_{}^{}}^{} 
\end{pmatrix}^{}_{}
  + \mathbf{Tr}[\Gamma_{}^{}]\right] 
\mathbb{P}_{}^{}[\bm{\Upsilon}_{}^{},t],\nonumber\\
 \end{eqnarray}
 where $\bm{\nabla}_{\bm{\Upsilon}_{}^{}}^{} = \begin{pmatrix} 
\frac{\partial}{\partial\gamma_{1}^{*}} & \cdots & 
\frac{\partial}{\partial\gamma_{M}^{*}} & 
\frac{\partial}{\partial\gamma_{1}^{}} 
& \cdots & \frac{\partial}{\partial\gamma_{M}^{}} \end{pmatrix}_{}^{T}$,
%%(complex derivatives here should be interpreted in the sense of Wiritinger 
%% calculus [{\color{red}\textbf{Ref.} $\bm{\cdots}$}]), 
$\Gamma_{}^{} = \sum_{\alpha = 1}^{N} \Gamma_{\alpha}^{}$ and the ($2 \times 
2$ block partitioned) complex symmetric matrix $\mathbb{H}$ is defined as, 
\begin{eqnarray}
\label{eq-15}
 &&\mathbb{H}_{}^{}=\nonumber\\
 &&\frac{1}{2}\sum_{\alpha=1}^{N} 
\mathbb{V}[\chi_{\alpha}^{},\lambda_{\alpha}^{}]_{}^{T} \left[ \sigma_{x}^{} 
\otimes \mathbf{\Gamma}_{\alpha}^{}+\left(I_{}^{}-\sigma_{z}^{}\right)\otimes 
\left(\mathbf{\Gamma}_{\alpha}^{} \mathbf{D}_{\alpha}^{}\right)\right] 
\mathbb{V}[\chi_{\alpha}^{},\lambda_{\alpha}^{}]_{}^{}\nonumber\\
\end{eqnarray}
with 
\begin{eqnarray}
\label{eq-16}
&&\mathbb{V}_{}^{}[\chi_{}^{},\lambda_{}^{}] =e_{}^{i \sigma_{z}^{} \otimes 
\mathbf{h}_{S}^{} \lambda_{}^{}}\times\nonumber\\
&&\left[I_{}^{} \otimes \cos[\frac{1}{2}\mathbf{h}_{S}^{}\chi_{}^{}] - 
\frac{1}{4}\left(5 \sigma_{x}^{} -3 i \sigma_{y}^{} 
\right)\otimes\left(\bm{\sigma}_{y}^{}\sin[\frac{1}{2}\mathbf{h}_{S}^{}\chi_{}^{
}]\right)\right].\nonumber\\
\end{eqnarray}

The parabolic partial differential equation, Eq. (\ref{eq-14}), can be 
analytically solved. A brief description of two methods that can be used to 
solve this class of equations is given in the 
appendix. It is to be noted that similar type of partial differential 
equations also appeared in the studies of heat current fluctuations through  
classical harmonic chains \cite{kundu2011large,dhar2015heat} and work 
statistics of driven classical harmonic oscillators subjected to thermal noise 
\cite{gupta2017stochastic,manikandan2017asymptotics}. Also, a related 
partial differential equation is encountered in the study of work statistics of 
degenerate parametric amplification process \cite{Yadalam2019}. 

Solution of Eq. (\ref{eq-14}) is given in terms of a Green function as, 
\begin{eqnarray}
\label{eq-17}  
\mathbb{P}_{}^{}[\bm{\Upsilon}_{}^{},t]&=&\int_{}^{}\mathcal{D}[\bm{\Upsilon}_{}
^{\prime}]\
\mathbb{G}_{}^{}[\bm{\Upsilon}_{}^{},t_{}^{}|\bm{\Upsilon}_{}^{\prime},0]\mathbb
{P}_{}^{}[\bm{\Upsilon}_{}^{\prime},0]
 \end{eqnarray}
with the Greens function given by,
\begin{widetext}
 \begin{eqnarray}
 \label{eq-18}
  \mathbb{G}_{}^{}[\bm{\Upsilon}_{}^{},t_{}^{}|\bm{\Upsilon}_{}^{\prime},0]&=&
  \frac{1}{\pi_{}^{M}}\frac{e^{\frac{1}{2} \mathbf{Tr}[\Gamma_{}^{}] 
t_{}^{}}_{}}{\sqrt{\mathbf{Det}\left[\mathbb{U}_{21}^{}(t)\bm{\sigma}_{x}^{}
\right]}}
  e^{-\frac{1}{2} \left\{ \bm{\Upsilon}_{}^{T}  
\left[\mathbb{U}_{12}^{}(t){\mathbb{U}_{22}^{}(t)}_{}^{-1}\right] 
\bm{\Upsilon}_{}^{}+\left[\bm{\Upsilon}_{}^{} -\mathbb{U}_{22}^{}(t) 
\bm{\Upsilon}_{}^{\prime} \right]_{}^{T} 
\left[\mathbb{U}_{21}^{}(t){\mathbb{U}_{22}^{}(t)}_{}^{T}\right]_{}^{-1} 
  \left[\bm{\Upsilon}_{}^{} -\mathbb{U}_{22}^{}(t) \bm{\Upsilon}_{}^{\prime} 
\right]_{}^{}\right\}}_{}.
  \end{eqnarray}
\end{widetext}
Here $\mathbb{U}_{xy}^{}(t)$ are $2 M \times 2 M$ matrices defined as the $2 
\times 2$ blocks of (block partitioned complex symplectic 
matrix) $\mathbb{U}_{}^{}(t)$, defined as, 
\begin{eqnarray}
\label{eq-19}
 \mathbb{U}_{}^{}(t) &=& \begin{pmatrix} \mathbb{U}_{11}^{}(t) & 
\mathbb{U}_{12}^{}(t) \\ \mathbb{U}_{21}^{}(t) & 
\mathbb{U}_{22}^{}(t)\end{pmatrix} = 
e_{}^{-\mathbb{H}_{}^{}\mathbf{\Sigma}_{}^{} t},
\end{eqnarray}
with the standard symplectic matrix, $\mathbf{\Sigma}_{}^{} = i \sigma_{y}^{} 
\otimes \mathbf{I}_{2 M \times 2 M}^{}$. 

The initial Wigner function of the system's initial state, the squeezed thermal 
state \cite{Wang2007}, is given as,
\begin{eqnarray}
\label{eq-20}
 \mathbb{P}_{}^{}[\bm{\Upsilon}_{}^{},0] &=& 
\frac{1}{\pi_{}^{M}}\frac{1}{\sqrt{\mathbf{Det}\left[\mathbf{D}_{S}^{}\bm{\sigma
}_{x}^{}\right]}}e_{}^{-\frac{1}{2}\bm{\Upsilon}_{}^{T}{\mathbf{D}_{S}^{}}_{}^{
-1}\bm{\Upsilon}_{}^{}}
\end{eqnarray}
with $\mathbf{D}_{S}^{} = -i \bm{\sigma}_{y}^{} e_{}^{-i \mathbf{S}_{S}^{} 
\bm{\sigma}_{y}^{}}\left[ n_{S}^{}(\mathbf{h}_{S}^{}) + \frac{1}{2} 
\mathbf{I}_{}^{}\right]e_{}^{i \mathbf{S}_{S}^{} 
\bm{\sigma}_{y}^{}}$, $\mathbf{S}_{S}^{} = Z_{S}^{} \bm{\sigma}_{z}^{} e_{}^{i 
\bm{\sigma}_{z}^{} \phi_{S}^{}}$ and $n_{S}^{}(x)=\left(e_{}^{\beta_{S}^{} 
x}-1\right)_{}^{-1}$. Using this in Eq. (\ref{eq-17}) and performing 
$\bm{\Upsilon}_{}^{\prime}$ Gaussian integral along with the use of identities 
derived from the symplectic property of $\mathbb{U}_{}^{}(t)$, i.e., 
${\mathbb{U}_{}^{}(t)}_{}^{T}\mathbf{\Sigma}_{}^{}\mathbb{U}_{}^{}(t) = 
\mathbf{\Sigma}_{}^{}$, an explicit form of the time-dependent Wigner function 
is obtained. This is given as,
\begin{eqnarray}
\label{eq-21}
 &&\mathbb{P}_{}^{}[\bm{\Upsilon}_{}^{},t] = \nonumber\\
 &&\frac{1}{\pi_{}^{M}} 
\frac{e_{}^{\frac{1}{2}\mathbf{Tr}\left[\Gamma_{}^{}\right] 
t}e_{}^{-\frac{1}{2}\bm{\Upsilon}_{}^{T}\left\{\left[\mathbb{U}_{11}^{}
(t)+\mathbb{U}_{12}^{}(t)\mathbf{D}_{S}^{}\right]\left[\mathbb{U}_{21}^{}
(t)+\mathbb{U}_{22}^{}(t)\mathbf{D}_{S}^{}\right]_{}^{-1}\right\}\bm{\Upsilon}_{
}^{}}}{\sqrt{\mathbf{Det}\left[\left[\mathbb{U}_{21}^{}(t)+\mathbb{U}_{22}^{}
(t)\mathbf{D}_{S}^{}\right]\bm{\sigma}_{x}^{}\right]}}.\nonumber\\
\end{eqnarray}
Using this in Eq. (\ref{eq-12}), and performing the Gaussian 
$\bm{\Upsilon}_{}^{}$ integral, the following expression is obtained,
\begin{eqnarray}
\label{eq-22}
 \tilde{\mathcal{Z}}[\bm{\chi}_{}^{},\bm{\lambda}_{}^{},t_{}^{}] &=& 
\frac{e_{}^{\frac{1}{2}\mathbf{Tr}\left[\Gamma_{}^{}\right] 
t}}{\sqrt{\mathbf{Det}\left[\mathbb{U}_{11}^{}(t)+\mathbb{U}_{12}^{}(t)\mathbf{D
}_{S}^{}\right]}}.
\end{eqnarray}
The above expression for 
$\tilde{\mathcal{Z}}[\bm{\chi}_{}^{},\bm{\lambda}_{}^{},t_{}^{}]$ can be 
considered as the dissipative generalization of the Levitov-Lesovik-Klich 
formula \cite{levitov1993charge,klich2003elementary}.

We note that in Ref. \cite{pigeon2015thermodynamics}, a method using 
phase-space quasi-probability functions, similar in spirit as discussed 
above, for computing long-time statistics of fluxes through quantum harmonic 
networks was developed. Noteworthy difference of our 
approach is that, it allows one to compute statistics in the transient regime. 
Furthermore, our approach is based on microscopic master equation and two-point 
measurement scheme, as opposed to Ref. \cite{pigeon2015thermodynamics}, which is 
based on counting of quantum jumps \cite{garrahan2010thermodynamics} of 
system described by a generic phenomenological master equations.

It is important to note that for $\bm{\chi}_{}^{} =\mathbf{0}_{}^{}$, 
$\mathbb{U}_{12}^{}(t)$ reduces to $2 M \times 2 M$ null matrix and 
$\mathbb{U}_{11}^{}(t)=e_{}^{\frac{1}{2}\left[I_{}^{} \otimes 
\Gamma_{}^{}\right]t_{}^{}}$, and thus Eq. (\ref{eq-22}) gives 
$\tilde{\mathcal{Z}}[\mathbf{0}_{}^{},\bm{\lambda}_{}^{},t_{}^{}]=1$. This 
indicates that $\mathcal{Z}[\bm{0}_{}^{},t_{}^{}]$ (defined in Eq. 
(\ref{eq-5})) is a divergent quantity. This divergence of 
$\mathcal{Z}[\bm{0}_{}^{},t_{}^{}]$ is not an artifact of the markov 
approximation used here. As already pointed out, it can be 
seen from the initial definition of $\mathcal{Z}[\bm{\chi}_{}^{},t_{}^{}]$ (Eq. 
(\ref{eq-5})) by using 
$\tilde{\mathcal{Z}}[\bm{0}_{}^{},\bm{\lambda}_{}^{},t_{}^{}]=1$ 
(can be seen by substituting $\bm{\chi}=0$ in Eq. (\ref{eq-6}) and using cyclic 
invariance of trace). Furthermore, 
it turns out that for the simple models discussed in the next 
section, $\mathcal{Z}[\bm{\chi}_{}^{},t_{}^{}]$ itself diverges as a result of 
markov approximation used here \cite{Yadalam2022}. For it to 
represent a meaningful moment generating function, we have to re-normalize it, 
so that the resultant probability function is normalized and meaningful. This 
renormalization can be achieved by dividing the value of 
$\mathcal{Z}[\bm{\chi}_{}^{},t_{}^{}]$ by $\mathcal{Z}[0,t_{}^{}]$. Since both 
these quantities diverge, this renormalization is performed after regularizing 
$\mathcal{Z}[\bm{\chi}_{}^{},t_{}^{}]$ by introducing a cutoff on the 
$\bm{\lambda}$ integral and taking the cutoff to infinity after division. This 
introduced cutoff, can be thought of as arising physically, by working with 
reservoirs with mode frequencies that are equally spaced with a small spacing 
($\bar{\epsilon}$) (for which initial projection can be implemented by 
$\bm{\lambda}$ integrals with an ultraviolet cutoff 
$|\bm{\lambda}_{k}|\leq \frac{\pi}{\bar{\epsilon}}$), which is sent to zero 
eventually. This renormalization is done case by case in the following. 

In the next section we apply the general results obtained in this section to 
two special cases, both with the single cavity mode coupled 
either to a single reservoir or to two reservoirs. 
\section{Application to simple models}
\label{sec4}
We now specialize to the case of a cavity with a single photon mode, i.e., we 
apply the results presented in the previous section to the case $M_{}^{}=1$. 
For this case $h_{S}^{}$, $\Gamma_{\alpha}^{}$, $I_{M \times M}^{}$ become 
scalars and $\mathbf{h}_{S}^{}$, $\mathbf{S}_{S}^{}$, $\mathbf{D}_{S}^{}$, 
$\mathbf{S}_{\alpha}^{}$ and $\mathbf{D}_{\alpha}^{}$ become $2 \times 2$ 
matrices, $\mathbb{U}_{}^{}(t)$ and $\mathbf{\Sigma}_{}^{}$ reduce to $4 \times 
4$ matrices and, hence, $\mathbb{U}_{xy}^{}(t)$ are $2 \times 2$ matrices. 
For later convenience, we also define $D_{\alpha}^{} = -i \sigma_{y}^{} 
e_{}^{-i S_{\alpha}^{} \sigma_{y}^{}}\left[ n_{\alpha}^{}(\epsilon_{}^{} 
\sigma_{z}^{}) + \frac{1}{2} I_{}^{}\right]e_{}^{i S_{\alpha}^{} 
\sigma_{y}^{}}$, $\epsilon_{}^{} = \epsilon_{1}^{}$, $S_{\alpha}^{} = 
Z_{\alpha}^{} \sigma_{z}^{} e_{}^{i \sigma_{z}^{} \phi_{\alpha}^{}}$ and 
$n_{\alpha}^{}(x)=\left(e_{}^{\beta_{\alpha}^{} x}-1\right)_{}^{-1}\equiv
n_{\alpha}^{}$ (for $\alpha = S,1, \cdots, N$). 

Below we consider two simple cases. First one is consisting of only one photon 
reservoir, while the second case is with two photon reservoirs.

\subsection{Single mode coupled to a single reservoir}
In this subsection, we present results for a model system consisting of a 
single photon mode cavity coupled to a single squeezed thermal photon 
reservoir, 
i.e., we further specialize to the case $N_{}^{}=1$. Using the explicit 
expressions for $\mathbb{U}_{11}^{}(t)$ and $\mathbb{U}_{12}^{}(t)$ in Eq. 
(\ref{eq-22}) with $\bm{\chi}_{}^{} = \chi_{1}^{}$ and $\bm{\lambda}_{}^{} = 
\lambda_{1}^{}$ gives,
\begin{eqnarray}
\label{eq-23}
 \tilde{\mathcal{Z}}[\chi_{1}^{},\lambda_{1}^{},t] &=& 
\displaystyle{\frac{e_{}^{\frac{\Gamma_{1}^{}t_{}^{}}{2}}}{\sqrt{\textbf{Det}
\left[\cosh[\frac{\Gamma_{1}^{}t_{}^{}}{2}]I_{}^{} 
+\sinh[\frac{\Gamma_{1}^{}t_{}^{}}{2}]\frac{\Xi_{1 
S}^{}[\chi_{1}^{},\lambda_{1}^{}]}{\frac{\Gamma_{1}^{}}{2}}\right]}}},
\nonumber\\
\end{eqnarray}
where $\Xi_{\alpha S}^{}[\chi_{\alpha}^{},\lambda_{\alpha}^{}]$ (here $\alpha = 
1$) is given as,
\begin{eqnarray}
\label{eq-24}
 &&\Xi_{\alpha S}^{}[\chi_{\alpha}^{},\lambda_{\alpha}^{}] = 
\frac{\Gamma_{\alpha}^{}}{2} I_{}^{}- \Gamma_{\alpha}^{} \times \nonumber\\
 && \left\{e_{}^{ i \epsilon_{}^{} \lambda_{\alpha}^{} 
\sigma_{z}^{}}\left[\sigma_{x}^{}D_{\alpha}^{}-\frac{1}{2}I_{}^{}\right]e_{}^{ 
- 
i \epsilon_{}^{} \lambda_{\alpha}^{} \sigma_{z}^{}}\left[ 
\sigma_{x}^{}D_{S}^{}+\frac{1}{2} I_{}^{}\right]\left(e_{}^{i \epsilon_{}^{} 
\chi_{\alpha}^{}}-1\right)\right. \nonumber\\
  && \left.+ e_{}^{ i \epsilon_{}^{} \lambda_{\alpha}^{} 
\sigma_{z}^{}}\left[\sigma_{x}^{}D_{\alpha}^{}+\frac{1}{2} I_{}^{}\right]e_{}^{ 
- i \epsilon_{}^{} \lambda_{\alpha}^{} 
\sigma_{z}^{}}\left[\sigma_{x}^{}D_{S}^{}-\frac{1}{2} 
I_{}^{}\right]\left(e_{}^{ 
-i \epsilon_{}^{} \chi_{\alpha}^{}}-1\right)\right\}.\nonumber\\
\end{eqnarray}

Substituting this expression for $\Xi_{1 S}^{}$ in Eq. (\ref{eq-23}) and 
upon simplification, we obtain, 
\begin{widetext}
\begin{eqnarray}
 \label{eq-s14}
 &&\tilde{\mathcal{Z}}[\chi_{1}^{},\lambda_{1}^{},t]=\nonumber\\
 &&e_{}^{\frac{\Gamma_{1}^{} 
t_{}^{}}{2}}\left\{\underset{x=\pm}{\prod}_{}^{}\left[\cosh[\frac{\Gamma_{1}^{}
t_{}}{2}]+\sinh[\frac{\Gamma_{1}^{}t_{}}{2}]\Lambda_{x}^{S}[\chi_{1}
^{}]\right] + 4
\left[1-e_{}^{-\Gamma_{1}^{}t_{}^{}}\right]\Delta_{1}^{}\Delta_{S}^{}\left[
\left(e_{}^{i \epsilon \chi_{1}^{}}-1\right)+\left(e_{}^{-i \epsilon 
\chi_{1}^{}}-1\right)\right]\sin_{}^{2}[\epsilon \lambda_{1}^{} + 
\frac{\phi_{1}^{}-\phi_{S}^{}}{2}]\right\}_{}^{-\frac{1}{2}},\nonumber\\
\end{eqnarray}
\end{widetext}
with
\begin{widetext}
 \begin{eqnarray}
  \label{eq-s15}
 \Lambda_{\pm}^{S}[\chi_{1}^{}] &=& 1 
-2\left\{\left[N_{1}^{}\pm\Delta_{1}^{}\right]\left[\left(1+N_{S}^{}
\right)\pm\Delta_{S}^{}\right]\left(e_{}^{i \epsilon 
\chi_{1}^{}}-1\right)+\left[\left(1+N_{1}^{}\right)\pm\Delta_{1}^{}\right]\left[
N_{S}^{}\pm\Delta_{S}^{}\right]\left(e_{}^{-i \epsilon 
\chi_{1}^{}}-1\right)\right\},
 \end{eqnarray}
\end{widetext}
with $N_{\alpha}^{}=\cosh_{}^{}[2 
Z_{\alpha}^{}]\left[n_{\alpha}^{}+\frac{1}{2}\right]-\frac{1}{2}$ 
and $\Delta_{\alpha}^{} = \sinh_{}^{}[2 
Z_{\alpha}^{}]\left[n_{\alpha}^{}+\frac{1}{2}\right]$.

The moment generating function for energy released from the system into the 
reservoir in time $t$ is then given by integrating over $\lambda_{1}^{}$ 
(defined in Eq. (\ref{eq-5})) as,
\begin{eqnarray}
\label{eq-25}
  \mathcal{Z}[\chi_{1}^{},t] &=& \frac{1}{2\pi}\int_{-\infty}^{+\infty} 
d_{}^{}\lambda_{1}^{}\ \tilde{\mathcal{Z}}[\chi_{1}^{},\lambda_{1}^{},t].
\end{eqnarray}
Since $\tilde{\mathcal{Z}}[\chi_{1}^{},\lambda_{1}^{},t]$ in Eq. 
(\ref{eq-s14}), 
is a periodic function of $\lambda_{1}^{}$ with period 
$\frac{2\pi}{\epsilon_{}^{}}$, i.e., 
$\tilde{\mathcal{Z}}[\chi_{1}^{},\lambda_{1}^{}+\frac{2\pi}{\epsilon_{}^{}},t]
=\tilde{\mathcal{Z}}[\chi_{1}^{},\lambda_{1}^{},t]$, 
$\mathcal{Z}[\chi_{1}^{},t]$ becomes divergent. To make sense of it as a 
moment generating function, we have to renormalize it. As discussed at the end 
of Sec. (\ref{sec3}), this is done by introducing 
a cutoff, $|\lambda_{1}^{}|\leq \frac{\pi}{\bar{\epsilon}_{}^{}}$ 
and re-normalizing $\mathcal{Z}[\chi_{1}^{},t]$ by $\mathcal{Z}[0,t]$ 
and taking the limit $\bar{\epsilon}_{}^{} \to 0$ as,
\begin{eqnarray}
\label{eq-26}
 \mathcal{Z}[\chi_{1}^{},t] &=& \lim_{\bar{\epsilon}_{}^{} \to 0}^{} 
\frac{\frac{1}{2\pi}\int_{-\frac{\pi}{\bar{\epsilon}_{}^{}}}^{+\frac{\pi}{\bar{
\epsilon}_{}^{}}} d_{}^{}\lambda_{1}^{}\ 
\tilde{\mathcal{Z}}[\chi_{1}^{},\lambda_{1}^{},t]}{\frac{1}{2\pi}\int_{-\frac{
\pi}{\bar{\epsilon}_{}^{}}}^{+\frac{\pi}{\bar{\epsilon}_{}^{}}} 
d_{}^{}\lambda_{1}^{}\ \tilde{\mathcal{Z}}[0,\lambda_{1}^{},t]}\nonumber\\
 &=& \frac{\epsilon_{}^{}}{2\pi} 
\int_{-\frac{\pi}{\epsilon_{}^{}}}^{+\frac{\pi}{\epsilon_{}^{}}} 
d_{}^{}\lambda_{1}^{}\ \tilde{\mathcal{Z}}[\chi_{1}^{},\lambda_{1}^{},t].
 \end{eqnarray}
To arrive at the second equality, we have used the periodic property of 
$\tilde{\mathcal{Z}}[\chi_{1}^{},\lambda_{1}^{},t]$ and 
$\tilde{\mathcal{Z}}[0,\lambda_{1}^{},t]=1$. The $\lambda_{1}^{}$ 
integral in the second equality can be analytically performed for 
$\tilde{\mathcal{Z}}[\chi_{1}^{},\lambda_{1}^{},t]$ given in Eq. 
(\ref{eq-s14}). This gives $\mathcal{Z}[\chi_{1}^{},t]$ in terms 
of complete elliptic function of first kind with the argument which is a 
complicated function of $\chi_{1}^{}$. Since this expression is not amenable to 
further analysis, we do not provide it here. However we note that, for the case 
when the initial states of system and reservoir are thermal, i.e., 
$Z_{1}^{}=Z_{S}^{}=0$, this expression for $\mathcal{Z}[\chi_{1}^{},t]$ 
agrees with the expressions previously reported in the literature 
\cite{Harbola2007,Novotny2015,Denzler2018} and the probability 
distribution function for the energy flow from the system into the reservoir 
satisfies the Jarzynski-Wojcik exchange fluctuation theorem 
\cite{Jarzynski2004Wojcik}.

Using $\mathcal{Z}[\chi_{1}^{},t]$, the cumulants of the energy flow from 
the system into the reservoir can be obtained. The average energy flow in time 
$t$ is given as, 
\begin{eqnarray}
\label{eq-27}
 &&\langle \Delta e_{1}^{} \rangle = \left(1-e_{}^{-\Gamma_{1}^{} 
t_{}^{}}\right)\left[N_{S}^{}-N_{1}^{}\right].
\end{eqnarray}
When the squeezing of the system and the reservoir are absent 
($Z_{1}^{}=Z_{S}^{}=0$), i.e., the system's initial state and the reservoir's 
state are thermal states, then $\langle \Delta e_{1}^{} \rangle = 
\left(1-e_{}^{-\Gamma_{1}^{} t_{}^{}}\right)\left[n_{S}^{}-n_{1}^{}\right]$. 
Comparing Eq. (\ref{eq-27}) with this allows us to define an effective 
(inverse) temperature in the presence of squeezing as, 
\begin{eqnarray}
\label{eq-28}
&&\tilde{\beta}_{\alpha}^{} =\frac{1}{\epsilon_{}^{}} 
\log\left[N_{\alpha}^{-1}+1\right].
\end{eqnarray}
As $N_{\alpha}^{}\geq n_{\alpha}^{}$ and $\log(x)$ is a 
monotonically increasing function, $\tilde{\beta}_{\alpha}^{-1}\geq 
\beta_{\alpha}^{-1}$. Hence, it is tempting to attribute the effect of 
squeezing to the enhancement of effective temperature of the reservoir. 
Using Eq. (\ref{eq-28}), the energy flow in the presence of squeezing can be 
expressed as, $\langle \Delta e_{1}^{} \rangle 
= \left(1-e_{}^{-\Gamma_{1}^{} 
t_{}^{}}\right)\left[\tilde{n}_{S}^{}-\tilde{n}_{1}^{}\right]$ with 
$\tilde{n}_{\alpha}^{} = \left(e_{}^{\tilde{\beta}_{\alpha}^{} \epsilon_{}^{}} 
- 1\right)_{}^{-1}$.\\

If it were true that the system's and reservoir's states could be described by 
thermal states with effective temperatures, then the energy flow from the 
system into the reservoir would satisfy the Jarzynski-Wojcik transient exchange 
fluctuation theorem with the effective temperature. However it turns out from 
the following discussion that the fluctuation theorem for the energy flow is 
absent for this system and hence, although the average energy flow can be 
described in terms of effective temperatures, this is not the case with its 
fluctuations. For instance, the second cumulant of the energy flow in time $t$ 
is given by, 
\begin{eqnarray}
 \label{eq-27b}
 \langle \Delta e_{1}^{2} \rangle_{}^{} &-& \langle \Delta e_{1}^{} 
\rangle_{}^{2}=(1-e_{}^{-\Gamma_{1}^{}t})_{}^{2}\left[(N_{S}^{}-N_{1}^{})_{}^{
2}+\Delta_{S}^{2} +\Delta_{1}^{2} 
\right]\nonumber\\&&+(1-e_{}^{-\Gamma_{1}^{}t})_{}^{} 
\left[N_{S}^{}(1+N_{1})+N_{1}^{}(1+N_{S}^{})\right],
\end{eqnarray}
cannot be expressed in terms of the effective temperature in the form, $ 
\langle \Delta e_{1}^{2} \rangle_{}^{} - \langle \Delta 
e_{1}^{}\rangle_{}^{2}=(1-e_{}^{-\Gamma_{1}^{}t})_{}^{2}\left[(\tilde{n}_{S}^{} 
-\tilde{n } _{ 1 } ^ { } )_{ } ^ {2}\right]+(1-e_{}^{-\Gamma_{1}^{}t})_{}^{} 
\left[\tilde{n}_{S}^{}(1+\tilde{n}_{1})+\tilde{n}_{1}^{}(1+\tilde{n}_{S}^{}
)\right]$, as obtained for the thermal case. This should be 
contrasted with a qubit coupled to a squeezed thermal reservoir, where it is 
possible to define an effective temperature such that the fluctuations of 
energy flow are same as that of the thermal case and the fluctuation theorem 
holds with an effective temperature \cite{Agarwalla2017}. 

Note that, in the long-time limit ($\Gamma_{1}^{}t_{}^{} \to \infty$), as the 
system reaches the same (``equilibrium'') state as that of the reservoir, the 
energy ceases to flow from the system into the reservoir and hence the energy 
flow and its fluctuations saturate to finite values. As a consequence, 
$\tilde{\mathcal{Z}}[\chi_{1}^{},\lambda_{1}^{},t]$, given in Eq. 
(\ref{eq-s14}), becomes independent of time, this indicates that the statistics 
of the energy flowing from the  system into the reservoir becomes independent 
of time. This is a generic feature of a finite system coupled to a single 
reservoir.\\

Owing to the periodicity, 
$\mathcal{Z}[\chi_{1}^{}+\frac{2\pi}{\epsilon_{}^{}},t] = 
\mathcal{Z}[\chi_{1}^{},t]$ (Eq. (\ref{eq-s14})), the probability function for 
the energy flow from system into the reservoir acquires a Dirac comb structure, 
i.e., 
$P[\Delta e_{1}^{},t] = \underset{n \in \mathbb{Z}_{}^{}}{\sum}_{}^{}p[n,t] 
\delta[\Delta e_{1}^{} - n \epsilon_{}^{}]$, with 
$p[n,t]=\frac{1}{2\pi} \int_{-\pi}^{+\pi} d_{}^{}\chi_{1}^{}\ 
\mathcal{Z}[\frac{\chi_{1}^{}}{\epsilon_{}^{}},t]e_{}^{i\chi_{1}^{} 
n_{}^{}}$. $p[n,t]$ is the probability of $n$-quanta of energy transferred from 
the system to the reservoir.\\ 

\begin{figure}[!tbh]
\centering
\includegraphics[width=8.6cm,height=5.2cm]
{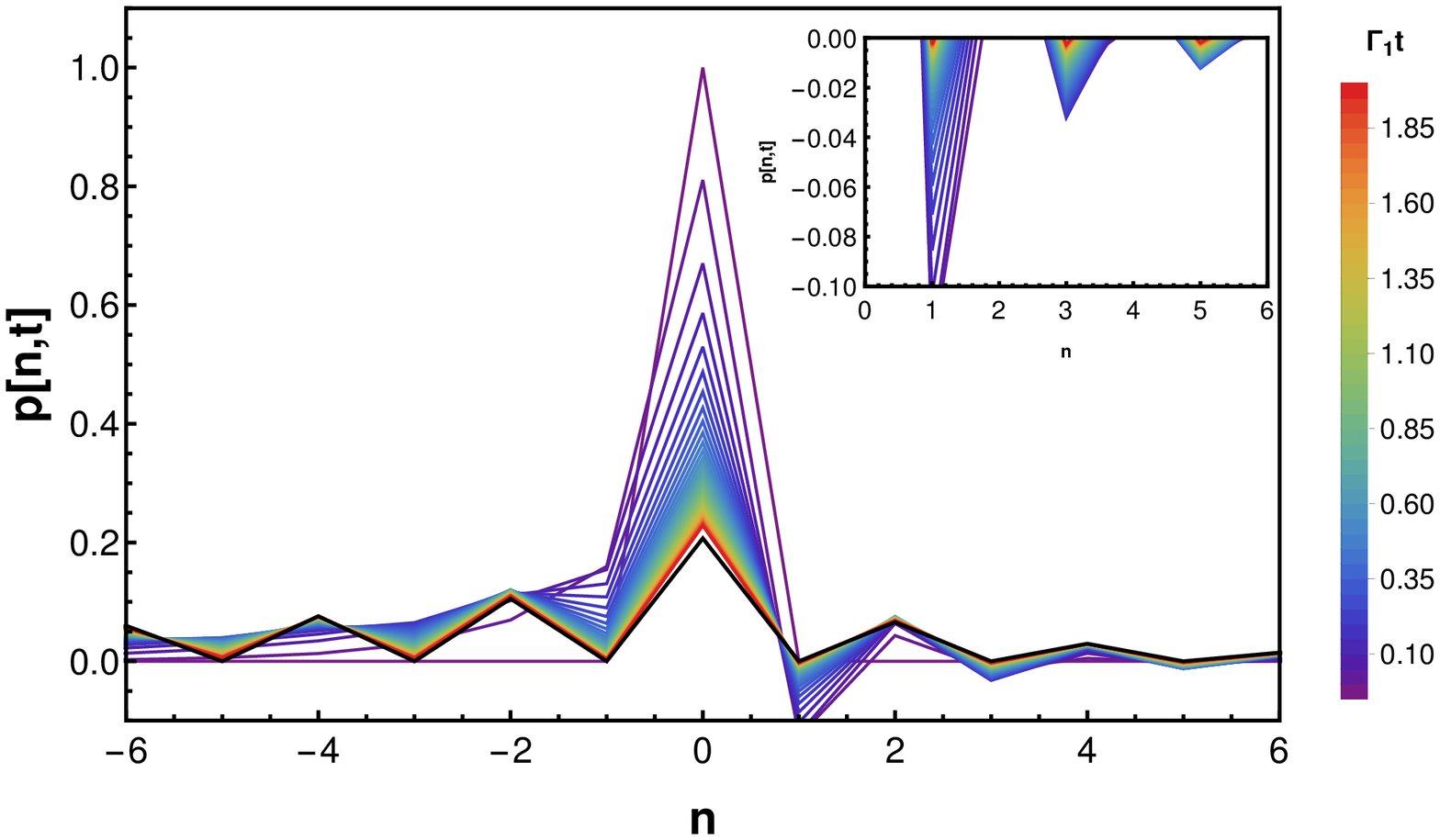}
\\~\\
\includegraphics[width=8.6cm,height=5.2cm]
{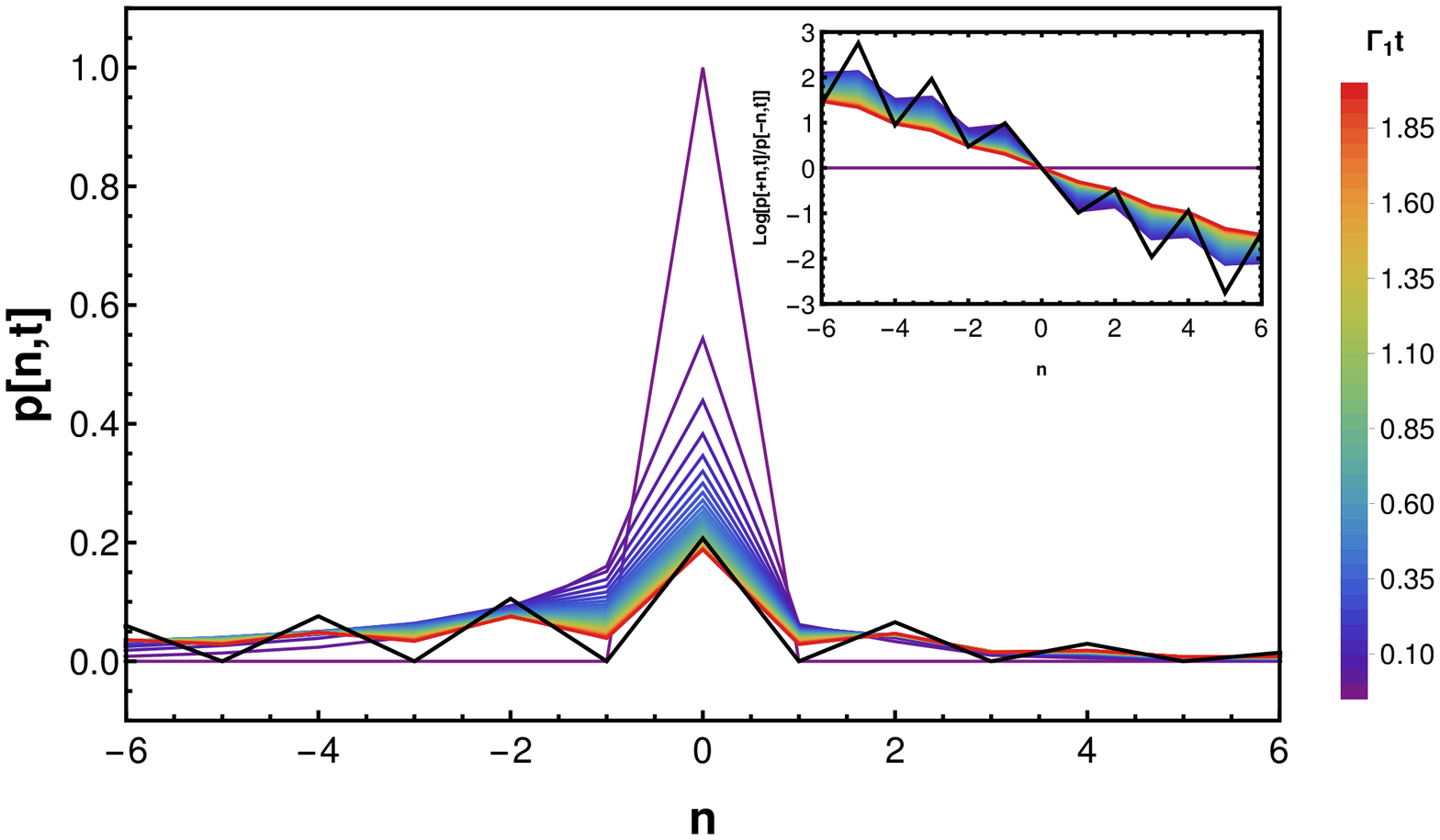}
\caption{Probability distribution function for number of quanta of energy 
released from the system into the reservoir in time $t$ for a range 
of $\Gamma_{1}^{} t$. The plot in the upper panel is 
obtained using $\tilde{\mathcal{Z}}[\chi_{1}^{},0,t]$ as the moment 
generating function (i.e., ignoring the non-commutativity of the initial 
projection and initial reservoir's state) with inset 
showing the region where $p[n,t]$ becomes negative. The plot in the lower 
panel is obtained using $\mathcal{Z}[\chi_{1}^{},t]$ (i.e., properly accounting 
for the initial projection) with the plots in the inset displaying 
$\log[p[n,t]/p[-n,t]]$ vs 
$n$. Black curves in both plots and their insets corresponds to $\Gamma_{1}^{} 
t \to \infty$. Parameters used are, 
$\beta_{1}^{}\epsilon_{}^{}=10.0$, $\beta_{S}^{}\epsilon_{}^{}=20.0$, 
$Z_{1}^{}=2.0$, $Z_{S}^{}=1.0$ and $\phi_{1}^{}-\phi_{S}^{}=\pi$.}
\label{single_probability}
\end{figure}

The $\lambda_{1}^{}$ dependence in 
$\tilde{\mathcal{Z}}[\chi_{1}^{},\lambda_{1}^{},t]$, which is integrated out to 
obtain the moment generating function in Eq. (\ref{eq-26}), contains 
information 
of the initial projective measurement on the reservoir. This 
$\lambda_{1}^{}$ integral has two important roles. Firstly, 
this makes $\mathcal{Z}[\chi_{1}^{},t]$ independent 
of the initial reservoir's and system's squeezing phases, $\phi_{1}^{}$ and 
$\phi_{S}^{}$, respectively. Hence the energy flow statistics is 
independent of these phases. This can be seen by performing a change of 
variables, $\lambda_{1}^{} \to \lambda_{1}^{} - 
\left(\frac{\phi_{1}^{}-\phi_{S}^{}}{2 \epsilon_{}^{}}\right)$, in the 
$\lambda_{1}^{}$ integral appearing in Eq. (\ref{eq-26}) along with the 
expression for $\tilde{\mathcal{Z}}[\chi_{1}^{},\lambda_{1}^{},t]$ given in Eq. 
(\ref{eq-s14}). Secondly, the $\lambda_{1}^{}$ integral is 
crucial for probability function, $p[n,t]$, to be meaningful. If we set 
$\lambda_{1}^{}=0$ 
to obtain, 
$\mathcal{Z}[\chi_{1}^{},t]=\tilde{\mathcal{Z}}[\chi_{1}^{},0,t]$, which is 
equivalent to the assumption that the initial energy projection 
commutes with the initial state of the reservoir, which is not the case here, 
we 
observe that the resulting moment generating function, 
$\mathcal{Z}[\chi_{1}^{},t]$, may lead to negative probabilities, $p[n,t]$, for 
certain events ($n$ values). This is evident from the plots shown in the upper  
panel (and the inset) of  Fig. (\ref{single_probability}), where negative 
probabilities are clearly evident for short time scales. The weight of negative 
probabilities decrease as time increases. In the long time limit ($\Gamma_{1}^{} 
t \to \infty$), it can be shown that, 
$\mathcal{Z}[\chi_{1}^{},t]=\tilde{\mathcal{Z}}[\chi_{1}^{},\lambda_{1}^{},t]
=\tilde{\mathcal{Z}}[\chi_{1}^{},0,t]$, making the long time statistics of 
energy flow independent of the initial energy projection, as 
it should be since the system reaches a well defined ``equilibrium'' state. 
More precisely, initial non-commutativity of the reservoirs density matrix with 
energy projective measurements does not affect the long time statistics. 
Figure in the lower panel uses the proper moment generating function, 
$\mathcal{Z}[\chi_{1}^{},t]$, obtained by accounting for the initial energy 
projection of the reservoir and gives the correct positive semi-definite 
distribution function $p[n,t]$ for all times. The negative probabilities 
observed previously in the statistics 
of charge flow between superconductors \cite{Bednorz2010,Shelankov2003} were  
attributed to the interference of transition amplitudes 
corresponding to different realizations (quantum trajectories) leading to the 
same energy change of the reservoir, but starting in different initial states 
\cite{Nazarov2003,Clerk2011,Hofer2016}. Finally, it is important to note 
that for the case, when either of the system's initial state or the 
reservoir's state is not squeezed, i.e.,  $Z_{S}^{}=0$ or $Z_{1}^{}=0$, 
$\tilde{\mathcal{Z}}[\chi_{1}^{},\lambda_{1}^{},t]$ given in Eq. 
(\ref{eq-s14}) becomes independent of $\lambda_{1}^{}$. Hence for this case, 
as expected, the statistics of energy flow is not affected by the 
non-commutative nature of the reservoir's density matrix with the initial 
projective measurement of the reservoir's energy.

Inset in the lower panel of Fig. (\ref{single_probability}) shows a 
nonlinear relationship between $\log[p[+n,t]/p[-n,t]]$ and $n$, indicating that 
the stochastic energy flow generically does not satisfy the Jarzynski-Wojcik 
exchange 
fluctuation theorem \cite{Jarzynski2004Wojcik} both at finite times as well as 
in the $\Gamma_{1}^{} t \to \infty$ limit. However for a special choice of 
parameters,
\begin{eqnarray}
 \label{eq-27c}
 Z_{1}^{}=\frac{\ln[1+2 n_{1}]}{2}\  \text{and}\  Z_{S}^{}=\frac{\ln[1+2 
n_{S}]}{2},
\end{eqnarray}
for which  $\Delta_{1}=N_{1}^{}$ and $\Delta_{S}^{}=N_{S}^{}$, 
the long-time ($\Gamma_{1}^{} t \to \infty$) moment generating function 
(obtained using Eq. (\ref{eq-s14}) in Eq. (\ref{eq-26})),
\begin{eqnarray}
 \label{eq-28a} 
&&\mathcal{Z}[\chi_{1}^{},\infty]=\{1-4\times\nonumber\\
&&\left[N_{1}^{}\left(1+2 
N_{S}^{}
\right)\left(e_{}^{i \epsilon 
\chi_{1}^{}}-1\right)+\left(1+2 
N_{1}^{}\right)N_{S}^{}\left(e_{}^{-i \epsilon 
\chi_{1}^{}}-1\right)\right]\}^{-\frac{1}{2}},\nonumber\\
\end{eqnarray}
exhibits Jarzynski-Wojcik exchange fluctuation theorem as a result of the 
Gallavotti-Cohen symmetry, 
$\mathcal{Z}[-\chi_{1}^{}-i \alpha_{1S},\infty]=\mathcal{Z}[\chi_{1
}^ {} , \infty ]$ \cite{Esposito2009}, with the affinity, 
$\alpha_{1S}=\ln\frac{N_{S}^{}\left(1+2 N_{1}^{}
\right)}{\left(1+ 2 N_{S}^{}\right) N_{1}^{}}$.

\subsection{Single mode coupled to two reservoirs}
In this subsection, we consider a single photon mode cavity coupled to two 
squeezed thermal photon reservoirs, i.e., we discuss the $N_{}^{}=2$ case. 
Unlike N=1 case, this allows to study fluctuations in a non-equilibrium 
steady-state. 

Using the explicit expressions for $\mathbb{U}_{11}^{}(t)$ and 
$\mathbb{U}_{12}^{}(t)$ in Eq. (\ref{eq-22}), we get an expression for 
$\tilde{\mathcal{Z}}[\bm{\chi}_{}^{},\bm{\lambda}_{}^{},t_{}^{}]$ given as,
\begin{eqnarray}
\label{eq-29}
 \tilde{\mathcal{Z}_{}^{}}[\bm{\chi}_{}^{},\bm{\lambda}_{}^{},t] &=& 
e_{}^{\left(\frac{\Gamma_{1}^{}+\Gamma_{2}^{}}{2}\right)t}\Bigg[1-\mathbb{X}_{--
}^{}[\bm{\chi}_{}^{},\bm{\lambda}_{}^{}]\nonumber\\
 &&+\displaystyle{\begin{bmatrix} 
\cosh(\Lambda_{-}^{}[\bm{\chi}_{}^{},\bm{\lambda}_{}^{}] t) \\ 
\frac{\sinh(\Lambda_{-}^{}[\bm{\chi}_{}^{},\bm{\lambda}_{}^{}] 
t)}{\Lambda_{-}^{}[\bm{\chi}_{}^{},\bm{\lambda}_{}^{}]} 
\end{bmatrix}_{}^{T}\mathbb{X}_{}^{}[\bm{\chi}_{}^{},\bm{\lambda}_{}^{}]
\begin{bmatrix} \cosh(\Lambda_{+}^{}[\bm{\chi}_{}^{},\bm{\lambda}_{}^{}] t) \\ 
\frac{\sinh(\Lambda_{+}^{}[\bm{\chi}_{}^{},\bm{\lambda}_{}^{}] 
t)}{\Lambda_{+}^{}[\bm{\chi}_{}^{},\bm{\lambda}_{}^{}]} 
\end{bmatrix}_{}^{}}\Bigg]_{}^{-\frac{1}{2}},\nonumber\\
\end{eqnarray}
where $\bm{\chi}_{}^ { } = \begin{pmatrix} \chi_{1}^{} & \chi_{2}^{} 
\end{pmatrix}_{}^{T}$, $\bm{\lambda}_{}^{} = \begin{pmatrix} \lambda_{1}^{} 
& \lambda_{2}^{} \end{pmatrix}_{}^{T}$ and
\begin{eqnarray}
\label{eq-30}
&&\Lambda_{\mp}^{}[\bm{\chi}_{}^{},\bm{\lambda}_{}^{}] =\nonumber\\
&&\sqrt{\left[\frac{\textbf{Tr}\left[\Xi_{12}^{}[\bm{\chi}_{}^{},\bm{\lambda}_{}
^{}]\right]}{2}\right]_{}^{} \pm 
\sqrt{\left[\frac{\textbf{Tr}\left[\Xi_{12}^{}[\bm{\chi}_{}^{},\bm{\lambda}_{}^{
}]\right]}{2}\right]_{}^{2} - 
\textbf{Det}\left[\Xi_{12}^{}[\bm{\chi}_{}^{},\bm{\lambda}_{}^{}]\right]}},
\nonumber\\
\end{eqnarray}
with
\begin{widetext}
\begin{eqnarray}
\label{eq-31}
 \Xi_{\alpha \alpha^\prime}^{}[\bm{\chi}_{}^{},\bm{\lambda}_{}^{}] = 
\left(\frac{\Gamma_{\alpha}^{}+\Gamma_{\alpha^\prime}^{}}{2}\right)_{}^{2}  
I_{}^{} - \Gamma_{\alpha}^{} \Gamma_{\alpha^\prime}^{} && \left\{e_{}^{i 
\epsilon_{}^{} \lambda_{\alpha}^{} 
\sigma_{z}^{}}\left[\sigma_{x}^{}D_{\alpha}^{}-\frac{1}{2} 
I_{}^{}\right]e_{}^{-i \epsilon_{}^{} 
\left(\lambda_{\alpha}^{}-\lambda_{\alpha^\prime}^{}\right) 
\sigma_{z}^{}}\left[\sigma_{x}^{}D_{\alpha^\prime}^{}+\frac{1}{2} 
I_{}^{}\right]e_{}^{-i \epsilon_{}^{} \lambda_{\alpha^\prime}^{} 
\sigma_{z}^{}}\left(e_{}^{i \epsilon_{}^{} 
\left(\chi_{\alpha}^{}-\chi_{\alpha^\prime}^{}\right)}-1\right)\right. 
\nonumber\\
  && \left.+ e_{}^{ i \epsilon_{}^{} \lambda_{\alpha}^{} 
\sigma_{z}^{}}\left[\sigma_{x}^{} D_{\alpha}^{}+\frac{1}{2} 
I_{}^{}\right]e_{}^{ 
- i \epsilon_{}^{} \left(\lambda_{\alpha}^{}-\lambda_{\alpha^\prime}^{}\right) 
\sigma_{z}^{}}\left[\sigma_{x}^{}D_{\alpha^\prime}^{}-\frac{1}{2} 
I_{}^{}\right]e_{}^{-i \epsilon_{}^{} \lambda_{\alpha^\prime}^{} 
\sigma_{z}^{}}\left(e_{}^{- i \epsilon_{}^{} 
\left(\chi_{\alpha}^{}-\chi_{\alpha^\prime}^{}\right)}-1\right)\right\}
\nonumber\\
\end{eqnarray}
\end{widetext}
and the explicit expressions for the matrix elements of the $2 \times 2$ 
matrix, 
$\mathbb{X}_{}^{}[\bm{\chi}_{}^{},\bm{\lambda}_{}^{}]$, are given in the 
appendix. 

Similar to the last section, 
$\tilde{\mathcal{Z}_{}^{}}[\bm{\chi}_{}^{},\bm{\lambda}_{}^{},t]$ is a 
periodic function of both $\lambda_{1}^{}$ and $\lambda_{2}^{}$ with period 
$\frac{2\pi}{\epsilon}$. Hence the joint moment generating function, 
$\mathcal{Z}[\bm{\chi}_{}^{},t]=
\int_{\bm{\lambda}_{}^{} \in 
\mathbb{R}_{}^{2}}^{}\frac{d_{}^{2}\bm{\lambda}_{}^{}}{(2\pi)_{}^{2}}\tilde{
\mathcal{Z}_{}^{}}[\bm{\chi}_{}^{},\bm{\lambda}_{}^{},t]$ diverges. To make 
$\mathcal{Z}[\bm{\chi}_{}^{},t]$ a proper moment generating function, we 
introduce two cutoffs in $\bm{\lambda}_{}^{}$ integrals, renormalize 
$\mathcal{Z}[\bm{\chi}_{}^{},t]$ by $\mathcal{Z}[\bm{0}_{}^{},t]$ and send 
the cutoffs to infinity to obtain the 
following expression, 
\begin{eqnarray}
\label{eq-32}
 \mathcal{Z}[\bm{\chi}_{}^{},t] &=& 
\left(\frac{\epsilon}{2\pi}\right)_{}^{2}\underset{\bm{\lambda}_{}^{} \in 
\left[-\frac{\pi}{\epsilon},+\frac{\pi}{\epsilon}\right]_{}^{2}}{\int}_{}^{}d_{}
^{2}\bm{\lambda}_{}^{}\ 
\tilde{\mathcal{Z}_{}^{}}[\bm{\chi}_{}^{},\bm{\lambda}_{}^{},t].
\end{eqnarray}
Furthermore, by doing the change of variables 
$\lambda_{1/2}^{} \to 
\lambda_{1/2}^{} - \left(\frac{\phi_{1/2}^{}-\phi_{S}^{}}{2 
\epsilon_{}^{}}\right)$ and using periodic property of 
$\tilde{\mathcal{Z}_{}^{}}[\bm{\chi}_{}^{},\bm{\lambda}_{}^{},t]$ with respect 
to $\lambda_{1/2}^{}$, it can be shown that the squeezing phases of the initial 
states of the system ($\phi_{S}^{}$) and both the reservoirs ($\phi_{1}^{}$ and 
$\phi_{2}^{}$) do not affect the statistics of the energy flow from the system 
into the reservoirs. 

For further analysis, it is convenient to consider the joint statistics of  
$\Delta e_{s}^{} = \left(\Delta e_{1}^{}+\Delta 
e_{2}^{}\right)$ and $\Delta e_{r}^{} = \frac{1}{2}\left(\Delta e_{1}^{}-\Delta 
e_{2}^{}\right)$, which, in the weak system-reservoir coupling limit considered 
in this work, can be interpreted as the net energy flow out of the system 
($\Delta e_{s}^{}$) and the net energy flow ($\Delta e_{r}^{}$) between the two 
reservoirs respectively. The joint moment generating function for these 
stochastic quantities can be obtained as 
$\bar{\mathcal{Z}}[\chi_{r}^{},\chi_{s}^{},t]=\mathcal{Z}[\bm{\chi}_{}^{},t]
\Big|_{\chi_{1/2}^{} \to \chi_{s}^{} \pm \frac{1}{2}\chi_{r}^{}}^{}$, where 
$\chi_{r}$ and $\chi_{s}$ are parameters conjugate to $\Delta e_{r}$ and 
$\Delta e_{s}$ respectively. 

The marginal moment generating function corresponding to $\Delta e_{s}^{}$, 
$\mathcal{Z}_{s}^{}[\chi_{s}^{},t]=\bar{\mathcal{Z}}[0,\chi_{s}^{},t]$ is 
obtained as, 

\begin{eqnarray}
\label{eq-33}
 \mathcal{Z}_{s}^{}[\chi_{s}^{},t] &=& 
\left(\frac{\epsilon}{2\pi}\right)_{}^{2}\underset{\bm{\lambda}_{}^{} \in 
\left[-\frac{\pi}{\epsilon},+\frac{\pi}{\epsilon}\right]_{}^{2}}{\int}_{}^{}d_{}
^{2}\bm{\lambda}_{}^{}\ 
\tilde{\mathcal{Z}}_{s}^{}[\chi_{s}^{},\bm{\lambda}_{}^{},t],
\end{eqnarray}
with
\begin{eqnarray}
\label{eq-34}
&&\tilde{\mathcal{Z}}_{s}^{}[\chi_{s}^{},\bm{\lambda}_{}^{},t] =\nonumber\\ 
&&\displaystyle{\frac{e_{}^{\left(\frac{\Gamma_{1}^{}+\Gamma_{2}^{}}{2}\right)t_
{}^{}}}{\sqrt{\textbf{Det}\left[\cosh[\left(\frac{\Gamma_{1}^{}+\Gamma_{2}^{}}{2
}\right)t_{}^{}]I_{}^{} 
+\sinh[\left(\frac{\Gamma_{1}^{}+\Gamma_{2}^{}}{2}\right)t_{}^{}]\frac{\Xi_{R 
S}^{}[\chi_{s}^{},\bm{\lambda}_{}^{}]}{\left(\frac{\Gamma_{1}^{}+\Gamma_{2}^{}}{
2}\right)}\right]}}},\nonumber\\
\end{eqnarray}
where  
\begin{widetext}
 \begin{eqnarray}
  \Xi_{R S}^{}[\chi_{s}^{},\bm{\lambda}_{}^{}] = 
\left[\sum_{\alpha=1}^{2}\frac{\Gamma_{\alpha}^{}}{2}\right] I_{}^{}- 
&&\left\{\sum_{\alpha=1}^{2}\Gamma_{\alpha}^{}\left[e_{}^
{ i \epsilon_{}^{} \lambda_{\alpha}^{} 
\sigma_{z}^{}}\left[\sigma_{x}^{}D_{\alpha}^{}-\frac{1}{2}I_{}^{}\right]e_{}^{ 
- 
i \epsilon_{}^{} \lambda_{\alpha}^{} \sigma_{z}^{}}\right]\left[ 
\sigma_{x}^{}D_{S}^{}+\frac{1}{2} I_{}^{}\right]\left(e_{}^{i \epsilon_{}^{} 
\chi_{s}^{}}-1\right)\right. \nonumber\\
  && \left.+ 
\sum_{\alpha=1}^{2}\Gamma_{\alpha
}^{}\left[e_{}^{ i \epsilon_{}^{} \lambda_{\alpha}^{} 
\sigma_{z}^{}}\left[\sigma_{x}^{}D_{\alpha}^{}+\frac{1}{2} I_{}^{}\right]e_{}^{ 
- i \epsilon_{}^{} \lambda_{\alpha}^{} 
\sigma_{z}^{}}\right]\left[\sigma_{x}^{}D_{S}^{}-\frac{1}{2} 
I_{}^{}\right]\left(e_{}^{ -i \epsilon_{}^{} 
\chi_{s}^{}}-1\right)\right\}.
\end{eqnarray}
\end{widetext}
The expression for 
$\tilde{\mathcal{Z}}_{s}^{}[\chi_{s}^{},\bm{\lambda}_{}^{},t]$ given above in  
Eq. (\ref{eq-34}), apart from its dependence on $\lambda_{1}^{}$ and 
$\lambda_{2}^{}$, has the similar mathematical structure as for the moment 
generating function  for energy transfer in presence of a single bath as given 
in Eq. (\ref{eq-23}). This 
indicates that the dynamical behavior of the statistics of the system's energy 
loss to reservoirs is similar to the case of single reservoir. Further, from 
Eq. (\ref{eq-34}), it is clear that 
$\lim_{t \to \infty}^{} \mathcal{Z}_{s}^{}[\chi_{s}^{},t]$ is finite, 
indicating that the statistics of $\Delta e_{s}^{}$ also becomes independent of 
time in the long time limit. This indicates that the fluctuations 
of energy flow out of the system saturate with time as the system reaches 
steady-state. From here onwards, we confine ourselves to the steady 
state and only discuss the statistics of the energy flow from the reservoir '2' 
into 
the reservoir '1' ($\Delta e_{r}^{}$), i.e., we only analyze the marginal 
distribution function $P[\Delta e_{r}^{},t]$ in the $t \to \infty$ limit.

In the long time limit ($t \to \infty$), moment generating function 
corresponding to the energy flow ($\Delta e_{r}^{}$), 
defined as 
$\mathcal{Z}_{r}^{}[\chi_{r}^{},t]=\bar{\mathcal{Z}}[\chi_{r}^{},0,t]$, is 
obtained by substituting the leading term of Eq. (\ref{eq-29}) in 
Eq. (\ref{eq-32}). This is given as, 
\begin{widetext}
\begin{eqnarray}
\label{eq-35s}
 &&\mathcal{Z}_{r}^{}[\chi_{r}^{},t] = 
\left(\frac{\epsilon}{2\pi}\right)_{}^{2}\underset{\bm{\lambda}_{}^{} \in 
\left[-\frac{\pi}{\epsilon},+\frac{\pi}{\epsilon}\right]_{}^{2}}{\int}_{}^{}d_{}
^{2}\bm{\lambda}_{}^{}\nonumber\\ 
&&2\left\{\left[\mathbb{X}_{--}^{}[\bm{\chi}_{}^{},\bm{\lambda}_{}^{}]+\frac{
\mathbb{X}_{-+}^{}[\bm{\chi}_{}^{},\bm{\lambda}_{}^{}]}{\Lambda_{+}^{}[\bm{\chi}
_{}^{},\bm{\lambda}_{}^{}]}+\frac{\mathbb{X}_{+-}^{}[\bm{\chi}_{}^{},\bm{\lambda
}_{}^{}]}{\Lambda_{-}^{}[\bm{\chi}_{}^{},\bm{\lambda}_{}^{}]}+\frac{\mathbb{X}_{
++}^{}[\bm{\chi}_{}^{},\bm{\lambda}_{}^{}]}{\Lambda_{+}^{}[\bm{\chi}_{}^{},\bm{
\lambda}_{}^{}]\Lambda_{-}^{}[\bm{\chi}_{}^{},\bm{\lambda}_{}^{}]}\right]_{}^{
-\frac{1}{2}}e_{}^{\left[\frac{\Gamma_{1}^{}+\Gamma_{2}^{}}{2} 
-\frac{\Lambda_{+}^{}[\bm{\chi}_{}^{},\bm{\lambda}_{}^{}]+\Lambda_{-}^{}[\bm{
\chi}_{}^{},\bm{\lambda}_{}^{}]}{2}\right] 
t_{}^{}}\right\}\Bigg|_{\chi_{1/2}^{} 
\to \pm \frac{1}{2}\chi_{r}^{}}^{}.
\end{eqnarray}
\end{widetext}
As noted already, the squeezing phases can be gauged to zero by shifting the 
integration variables $\bm{\lambda}$ in Eq. (\ref{eq-35s}), and hence we can 
set, $\phi_{S}^{}=\phi_{1}^{}=\phi_{2}^{}=0$.\\ 

For performing $\bm{\lambda}_{}^{}$ integrals, it is convenient to change 
the integration variables to $\lambda_{}^{}=\lambda_{1}^{}-\lambda_{2}^{}$ and 
$\bar{\lambda}_{}^{} = \frac{\lambda_{1}^{}+\lambda_{2}^{}}{2}$. Although 
$\Lambda_{\pm}^{}[\bm{\chi}_{}^{},\bm{\lambda}_{}^{}]$ depends only on 
$\lambda_{}^{}$ (this can be seen from Eq. (\ref{eq-30}) along with Eq. 
(\ref{eq-31})),  $\mathbb{X}_{\pm\pm}^{}[\bm{\chi}_{}^{},\bm{\lambda}_{}^{}]$ 
depend on both $\lambda_{}^{}$ and $\bar{\lambda}_{}^{}$. However, when the 
system's initial state is not squeezed, i.e., $Z_{S}^{}=0$, 
$\mathbb{X}_{\pm\pm}^{}[\bm{\chi}_{}^{},\bm{\lambda}_{}^{}]$ becomes 
independent of $\bar{\lambda}_{}^{}$. This is because the simultaneous  
measurements of both the reservoirs energies (in the weak coupling limit) is 
equivalent to measuring the system's energy and the difference of energies of 
the two reservoirs. And the $\bar{\lambda}_{}^{}$ dependence, which accounts 
for the non-commutativity of initial system's energy measurement with initial 
system's density matrix, drops out as system's initial state commutes with 
the initial energy projective measurement for this case. We focus on the 
statistics at steady-state where the system's initial state does not play a 
role. Therefore, for simplification purpose, we consider the case where the 
system's initial state is a thermal state. For this case, $\bar{\lambda}_{}^{}$ 
in Eq. (\ref{eq-35s}) can be integrated out, leaving only the $\lambda_{}^{}$ 
integral behind, which, in the long-time limit, is performed in the saddle point 
approximation. Saddle point of the exponent in Eq. (\ref{eq-35s}) is found at 
$\lambda_{}^{}=0$. This finally gives the steady-state scaled cumulant 
generating function, 
\begin{eqnarray}
\label{eq-35}
 \mathcal{F}[\chi_{r}^{}]&=&\lim_{t \to 
\infty}^{}\frac{\ln\mathcal{Z}_{r}^{}[\chi_{r}^{},t]}{t}=\frac{ 
\Gamma_{1}^{}+\Gamma_{2}^{}}{2}\sum_{x=\pm}^{}\left[\frac{1-\Lambda_{x}^{12}[
\chi_ {r}^{}]}{2}\right], \nonumber\\
\end{eqnarray}
with 
\begin{widetext}
 \begin{eqnarray}
 \label{eq-36}
 \Lambda_{\pm}^{12}[\chi_{r}^{}] &=& 
\sqrt{1 - 
\mathbb{T}\left\{\left[N_{1}^{}\pm\Delta_{1}^{}\right]\left[
\left(1+N_{2}^{}\right)\pm\Delta_{2}^{}\right]\left(e_{}^{i \epsilon 
\chi_{r}^{}}-1\right)+\left[\left(1+N_{1}^{}\right)\pm\Delta_{1}^{}\right]\left[
N_{2}^{}\pm\Delta_{2}^{}\right]\left(e_{}^{-i \epsilon 
\chi_{r}^{}}-1\right)\right\}}
 \end{eqnarray}
\end{widetext}
where $N_{X}^{}=\cosh_{}^{}[2 
Z_{X}^{}]\left[n_{X}^{}+\frac{1}{2}\right]-\frac{1}{2}$, $\Delta_{X}^{} = 
\sinh_{}^{}[2 Z_{X}^{}]\left[n_{X}^{}+\frac{1}{2}\right]$ with 
$n_{X}=\left(e^{\beta_{X}\epsilon}-1\right)^{-1}$ ($X=1,2$) and 
$\mathbb{T}=\frac{4\ \Gamma_{1}^{} 
\Gamma_{2}^{}}{\left(\Gamma_{1}^{}+\Gamma_{2}^{}\right)_{}^{2}}$.

The statistics of energy flux flowing between the two reservoirs can 
be computed using the above scaled cumulant generating function. 
The steady-state average flux is obtained as $\lim_{t \to \infty}\frac{\langle 
\Delta 
e_{r}\rangle}{t}=\frac{\Gamma_{1}\Gamma_{2}}{\Gamma_{1}+\Gamma_{2}}\left[N_{2}
-N_{1}\right]$. For $N_{1}=N_{2}=N_{}$, the energy flux 
between the reservoirs vanishes, however, it turns out that the probability 
function is not symmetric (i.e., skewed) around the origin ($n=0$), as the 
third cumulant, $\lim_{t \to \infty}\frac{\langle \Delta 
e_{r}^{3}\rangle_{c}}{t}=6 
\frac{\Gamma_{1}^{2}\Gamma_{2}^{2}}{\left(\Gamma_{1}^{ }+\Gamma_{2}^{}\right)^{3
}}\left[1+2 N_{}\right]\left[\Delta_{2}^{2}-\Delta_{1}^{2}\right]$, is nonzero. 
Hence according the two-point measurement scheme analysis, two squeezed 
thermal reservoirs can be considered at mutual equilibrium if their 
temperatures and squeezing amplitudes are same, although their phases may be 
different.\\

The marginal distribution function for the energy flow between reservoirs is 
then given as, 
\begin{eqnarray}
\label{eq-37}
 P[\Delta e_{r}^{},t] &=& 
\int_{-\infty}^{+\infty}\frac{d 
\chi_{r}^{}}{2\pi}\ 
 e_{}^{\mathcal{F}[\chi_{r}^{}] t +i \chi_{r}^{} \Delta 
e_{r}^{}}\nonumber\\
&=& \underset{n \in 
\mathbb{Z}_{}^{}}{\sum}_{}^{}p[n,t] 
\delta[\Delta e_{1}^{} - n \epsilon_{}^{}],\nonumber
\end{eqnarray} 
with $p[n,t]=\frac{1}{2\pi} 
\int_{-\pi}^{+\pi} d_{}^{}\chi_{r}^{}\ 
e_{}^{\mathcal{F}[\frac{\chi_{r}^{}}{\epsilon}] t_{}^{}+i\chi_{r}^{} 
n_{}^{}}$. The second equality in the above equation is a result of the 
periodicity of 
$\mathcal{F}[\chi_{r}^{}+\frac{2\pi}{\epsilon}]=\mathcal{F}[\chi_{r}^{}]$. 

In the long-time limit, we can define a large-deviation rate function 
$J[\frac{n}{t}]=-\lim_{t \to \infty}^{}\frac{p[n,t]}{t}$, such that $p[n,t] 
\overset{t \to \infty}{\approx} e_{}^{-J[\frac{n}{t}] t}$ 
\cite{Touchette2009,Esposito2009}.\\

\begin{figure}[!tbh]
\centering
\includegraphics[width=8.6cm,height=5.2cm]
{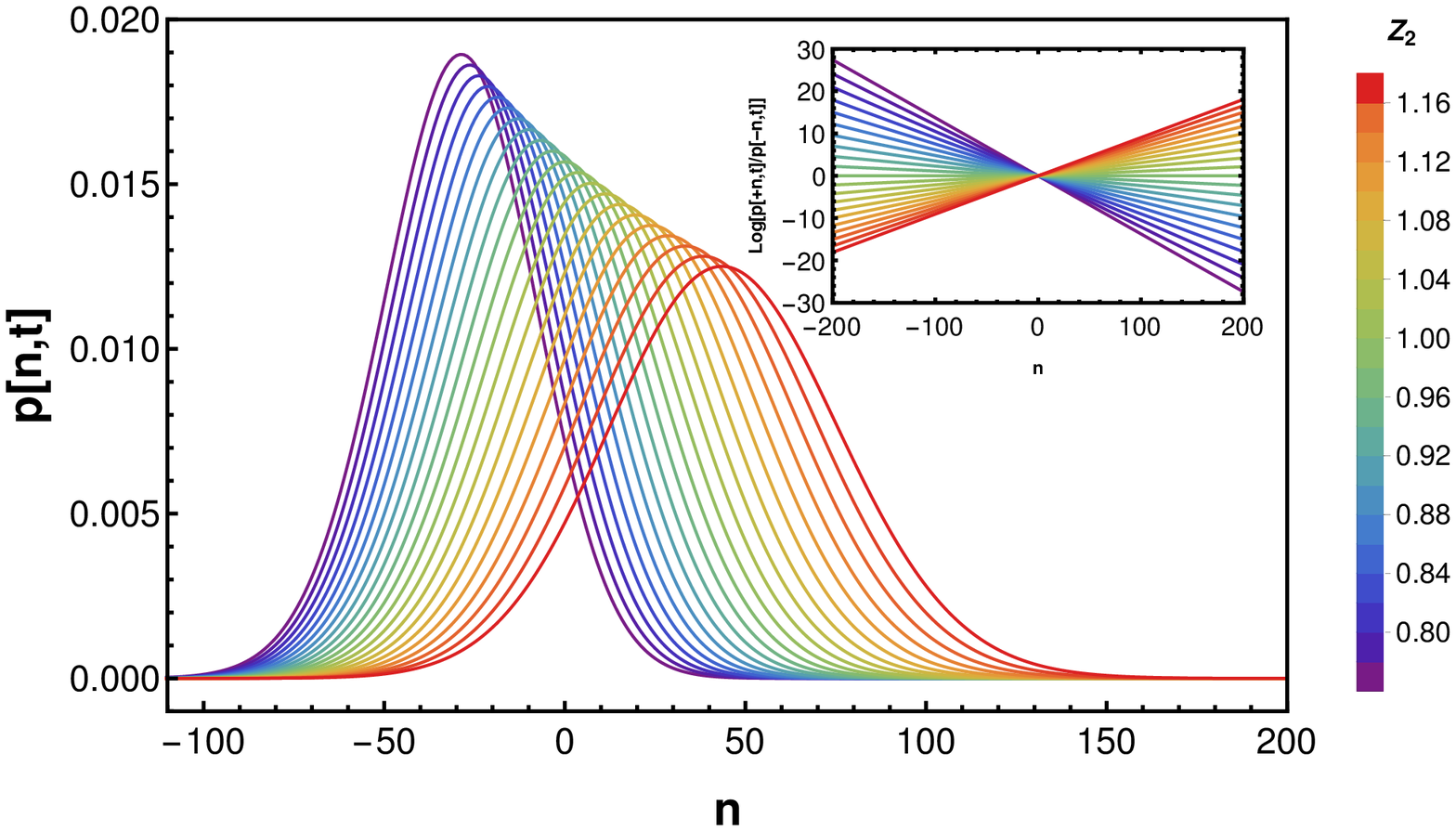}
\\~\\
\includegraphics[width=8.6cm,height=5.2cm]
{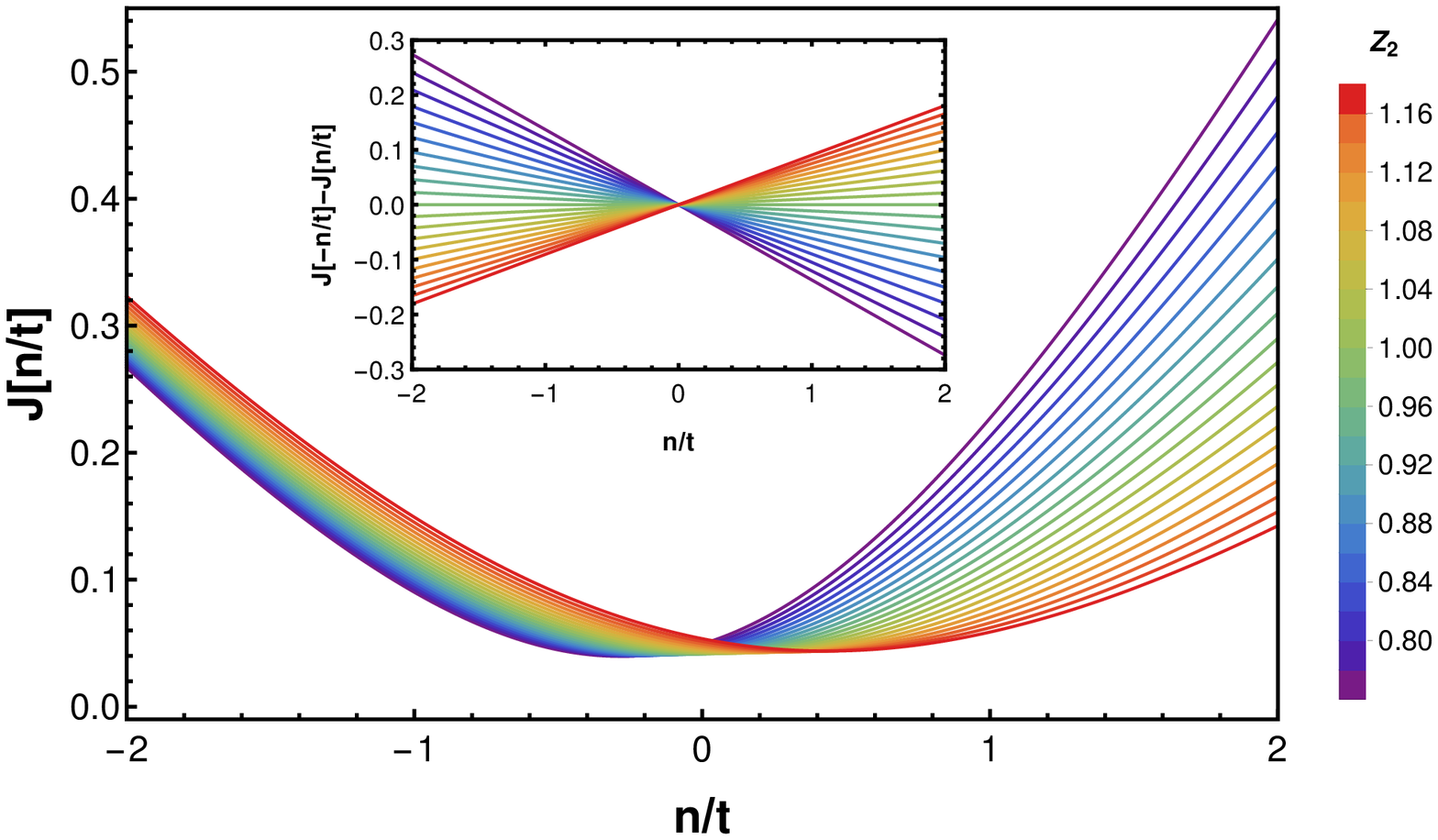}
\caption{(Upper panel) Marginal probability distribution function ($p[n,t]$) 
and (lower panel) the corresponding large deviation rate function ($J[n/t]$) 
for the number of quanta of energy exchanged between the reservoirs 
for a range of $Z_{2}^{}$ values in steady-state. Parameters used are, 
$\left(\frac{\Gamma_{1}^{}+\Gamma_{2}}{2}\right) 
t=100.0$, $\mathbb{T}=1$, 
$\beta_{1}^{}\epsilon_{}^{}=\beta_{2}^{}\epsilon_{}^{}=100.0$ and 
$Z_{1}^{}=1.0$. Inset (upper panel) shows the linearity of 
$\log\left[\frac{p[+n,t]}{p[-n,t]}\right]$ vs $n$ and (lower 
panel) the linearity of $J[-\frac{n}{t}]-J[\frac{n}{t}]$ vs $\frac{n}{t}$ 
($t \equiv \left(\frac{\Gamma_{1}^{}+\Gamma_{2}}{2}\right)_{}^{} t$).}
\label{double_probability}
\end{figure}

The marginal probability function and the corresponding rate function for the 
energy flow between reservoirs in the long-time limit are plotted in the upper 
and the lower panels of Fig. (\ref{double_probability}) respectively. The 
insets 
of these plots show respectively $\ln \frac{p[+n,t]}{p[-n,t]}$ vs $n$ 
and $J[-\frac{n}{t}]-J[\frac{n}{t}]$ vs $\frac{n}{t}$, which are both linear 
functions indicating the presence of Gallavotti-Cohen symmetry in 
$\mathcal{F}[\chi_{r}^{}]$ and steady-state fluctuation theorem for the 
marginal probability ($p[n,t]$). We were not able to 
identify the analytical form for the thermodynamic affinity due to the 
complexity of the steady-state cumulant generating function, Eq. (\ref{eq-35}). 
However, we note that, for a special set of parameters, $Z_{1}^{}=\frac{\ln[1+2 
n_{1}]}{2}$ and $Z_{2}^{}=\frac{\ln[1+2 n_{2}]}{2}$, such that  
$\Delta_{1}=N_{1}^{}$ and $\Delta_{2}^{}=N_{2}^{}$ (hence 
$\Lambda_{-}^{}[\chi_{r}^{}]=1$), 
a thermodynamic affinity, $\alpha_{12}=\frac{N_{2} (1+2 N_{1})}{(1+2 N_{2}) 
N_{1}}$ can be identified which determines the long time fluctuation 
theorem as a result of the Gallavotti-Cohen symmetry, 
$\mathcal{F}[-\chi_{r}^{}-i \alpha_{12}]=\mathcal{F}[\chi_{r}^ {}]$ 
\cite{Esposito2009}. Our numerical 
calculations indicate that the affinity is generically not an universal 
function of the reservoirs parameters (temperatures and squeezing amplitudes),  
although independent of the system-reservoir couplings, it depends also on the 
cavity mode frequency, which is a system-specific parameter. This is also 
evident from the above analytically identified affinity for the special set of 
parameters.

Thus unlike in the single reservoir case, where the long-time fluctuation 
theorem was recovered only for a special set of parameters, in the two 
reservoir case, the steady-state fluctuation theorem (with a non-universal 
affinity) is satisfied for all parameter values. 
\section{Conclusion}
A formalism for the analytical computation of  the full counting statistics of 
energy exchanged between a cavity weakly coupled to an arbitrary number of 
squeezed thermal photon reservoirs within two-point measurement scheme is 
developed. The crucial result of the formalism is Eq. (\ref{eq-22}) for the 
moment-generating function, which can be considered as the dissipative 
generalization of the Levitov-Lesovik-Klich formula. This formula is applied to 
two model systems, single mode cavity in contact with a single squeezed thermal 
reservoir and to two squeezed thermal reservoirs. It is found that the careful 
treatment of the initial projective measurement is necessary for getting 
physically meaningful probabilities for energy transport statistics at short 
times, although irrelevant for long-time scales. Generically, the full 
distribution function for the energy transfers is found not to satisfy transient 
fluctuation theorems. For the single reservoir case, a special parameter regime 
given by Eq. (\ref{eq-27c}) for which a steady-state fluctuation theorem  
emerges is identified. Contrary to this, for the two reservoir case, 
steady-state fluctuation theorem with a non-universal affinity is found to be 
valid always.  Furthermore, the analysis of the cumulants indicates that it is 
generically not possible to describe squeezed thermal reservoirs with an 
effective temperature, and two-squeezed thermal reservoirs cannot be considered 
as at equilibrium even if there is no energy flux between them and can be 
considered at equilibrium only if their temperatures and squeezing amplitudes 
are same. 
 \section*{Acknowledgements}
HKY acknowledges the hospitality of Indian Institute of Science  (India), where 
a major part of the work is carried out, and  International 
Centre for Theoretical Sciences (India) and \'Ecole Normale Sup\'erieure 
(France), where the final drafting of the work is done. HKY is also grateful 
for the support from the project 6004-1 of the Indo-French Centre for the 
Promotion of Advanced Research (IFCPAR). UH acknowledges support from Science 
and Engineering Research Board, India under the grant CRG/2020/001110.
BKA acknowledges the MATRICS grant MTR/2020/000472 from SERB, Government of 
India. BKA also thanks the Shastri Indo-Canadian Institute for providing 
financial support for this research work in the form of a Shastri Institutional 
Collaborative Research Grant (SICRG).

\section{Appendix}
\subsection{Methods for solving PDEs in Eq. (\ref{eq-14})}
In this section we provide brief sketch of two methods to solve the parabolic 
partial differential equation of the form encountered in the main text.
 \subsubsection{Method I}
 In this subsection we sketch a way to solve parabolic partial differential 
equations of the form,
\begin{eqnarray}
  \label{eq-s1}
   \frac{\partial}{\partial 
t}\mathbb{P}_{}^{}[\bm{\Upsilon}_{}^{},t]&=&\frac{1}{2}
  \left[ \begin{pmatrix} \bm{\Upsilon}_{}^{} \\ 
\bm{\nabla}_{\bm{\Upsilon}_{}^{}}^{} \end{pmatrix}^{T}_{}
  \mathbb{H}_{}^{}
  \begin{pmatrix} \bm{\Upsilon}_{}^{} \\ \bm{\nabla}_{\bm{\Upsilon}_{}^{}}^{} 
\end{pmatrix}^{}_{}
  + \mathbf{Tr}[\Gamma_{}^{}]\right] 
\mathbb{P}_{}^{}[\bm{\Upsilon}_{}^{},t],\nonumber\\
 \end{eqnarray}
with the initial condition 
$\mathbb{P}_{}^{}[\bm{\Upsilon}_{}^{},t]|_{t=0}^{}=\mathbb{P}_{}^{}[\bm{\Upsilon
}_{}^{},0]$. Here $\bm{\Upsilon}_{}^{} = \begin{pmatrix} \gamma_{1}^{*} & \cdots 
& \gamma_{M}^{*} & \gamma_{1}^{} & \cdots & \gamma_{M}^{} \end{pmatrix}_{}^{T}$, 
$\bm{\nabla}_{\bm{\Upsilon}_{}^{}}^{} = \begin{pmatrix} 
\frac{\partial}{\partial\gamma_{1}^{*}} & \cdots & 
\frac{\partial}{\partial\gamma_{M}^{*}} & \frac{\partial}{\partial\gamma_{1}^{}} 
& \cdots & \frac{\partial}{\partial\gamma_{M}^{}} \end{pmatrix}_{}^{T}$ 
%(complex derivatives here should be interpreted in the sense of Wiritinger 
%calculus) 
and $\mathbb{H}_{}^{} = 
\begin{pmatrix} \mathbb{H}_{11}^{} & \mathbb{H}_{12}^{} \\ \mathbb{H}_{21}^{} & 
\mathbb{H}_{22}^{} \end{pmatrix}$ is a $2 \times 2$ block partitioned $4 M 
\times 4 M$ complex symmetric matrix independent of $\bm{\Upsilon}_{}^{}$ and 
$t_{}^{}$. 
If $\mathbb{H}_{11}^{} = \mathbf{O}_{2 M \times 2 M}^{}$, the above equation is 
of the standard Ornstein-Uhlenbeck form, 
whose solution can be found in the Fourier domain by using method of 
characteristics \cite{Wang1945,VanKampen1992,Gardiner1994,Risken1996,
Carmichael2009}. 
For $\mathbb{H}_{11}^{} \neq \mathbf{O}_{2 M \times 2 M}^{}$, the quadratic 
term 
in the above equation can be eliminated using the transformation 
\cite{Polyanin2006}, 
\begin{eqnarray}
 \label{eq-s2}
\mathbb{P}_{}^{}[\bm{\Upsilon}_{}^{},t]=e_{}^{\frac{1}{2}\left[\bm{\Upsilon}_{}^
{T}\mathbb{R}_{}^{}(t)\bm{\Upsilon}_{}^{}+\mathbf{Tr}[\Gamma_{}^{}] 
t\right]}\bar{\mathbb{P}}_{}^{}[\bm{\Upsilon}_{}^{},t]
\end{eqnarray}
where without loss of generality, we can assume 
$\mathbb{R}_{}^{T}(t)=\mathbb{R}_{}^{}(t)$ and the requirement that 
$\mathbb{R}_{}^{}(t)$ satisfies the following Riccati matrix differential 
equation \cite{Reid1946,Reid1972,Kuvcera1973,Darling1997,Dahl2006,Kachalov2001,
Zelikin2013},
\begin{equation}
 \label{eq-s3}
 \frac{d}{dt}\mathbb{R}_{}^{}(t)=\mathbb{R}_{}^{}(t) \mathbb{H}_{22}^{} 
\mathbb{R}_{}^{}(t) + \mathbb{H}_{12}^{}\mathbb{R}_{}^{}(t) + 
\mathbb{R}_{}^{}(t) \mathbb{H}_{21}^{} + \mathbb{H}_{11}^{} 
\end{equation}
with the initial condition $\mathbb{R}_{}^{}(t)|_{t=0}^{}=\mathbf{O}_{2 M 
\times 
2 M}^{}$. 
The solution of this Riccati matrix differential equation is given as,
\begin{equation}
 \label{eq-s4}
 \mathbb{R}_{}^{}(t)=-\mathbb{U}_{12}^{}(t){\mathbb{U}_{22}(t)}_{}^{-1}
\end{equation}
where $\mathbb{U}_{xy}^{}(t)$ are blocks of $2 \times 2$ block partitioned $4 M 
\times 4 M$ complex symplectic matrix $\mathbb{U}_{}^{}(t)$ defined as,
\begin{equation} 
 \label{eq-s5}
 \mathbb{U}_{}^{}(t) = \begin{pmatrix} \mathbb{U}_{11}^{}(t) & 
\mathbb{U}_{12}^{}(t) \\ \mathbb{U}_{21}^{}(t) & 
\mathbb{U}_{22}^{}(t)\end{pmatrix} = 
e_{}^{-\mathbb{H}_{}^{}\mathbf{\Sigma}_{}^{} t},
\end{equation}
with the $4M \times 4M$ standard symplectic matrix defined as 
$\mathbf{\Sigma}_{}^{} = i \sigma_{y}^{} \otimes \mathbf{I}_{2 M \times 2 
M}^{}$. 
Using the symplectic property, 
$\mathbb{U}_{}^{T}(t)\mathbf{\Sigma}_{}^{}\mathbb{U}^{}(t)=\mathbf{\Sigma}_{}^{}
$, and the equation $\frac{d}{dt}\mathbb{U}_{}^{}(t) = - 
\mathbb{H}_{}^{}\mathbf{\Sigma}_{}^{} \mathbb{U}_{}^{}(t)$ (with 
$\mathbb{U}_{}^{}(t)|_{t=0}^{}=\mathbf{I}_{4 M \times 4 M}^{}$), 
$\bar{\mathbb{P}}_{}^{}[\bm{\Upsilon}_{}^{},t]$ is shown to satisfy the 
following parabolic partial differential equation of Ornstein-Uhlenbeck type, 
\begin{eqnarray}
  \label{eq-s6}
&&\frac{\partial}{\partial t}\bar{\mathbb{P}}_{}^{}[\bm{\Upsilon}_{}^{},t]
=\nonumber\\
   &&\frac{1}{2}\left[\begin{pmatrix} \bm{\Upsilon}_{}^{} \\ 
\bm{\nabla}_{\bm{\Upsilon}_{}^{}}^{} \end{pmatrix}^{T}_{} \begin{pmatrix} 
\mathbf{O}_{}^{} & \mathbb{H}_{12}^{} + \mathbb{R}_{}^{}(t) \mathbb{H}_{22}^{} 
\\ \mathbb{H}_{21}^{} + \mathbb{H}_{22}^{} \mathbb{R}_{}^{}(t) & 
\mathbb{H}_{22}^{} \end{pmatrix} \begin{pmatrix} \bm{\Upsilon}_{}^{} \\ 
\bm{\nabla}_{\bm{\Upsilon}_{}^{}}^{} 
\end{pmatrix}^{}_{}\right]\bar{\mathbb{P}}_{}^{}[\bm{\Upsilon}_{}^{},t],
\nonumber\\
 \end{eqnarray}
where $\mathbf{O}_{}^{}$ is the $2 M \times 2 M$ dimensional null matrix.This 
equation can be solved in Fourier domain using method of characteristics 
\cite{Wang1945,VanKampen1992,Gardiner1994,Risken1996,Carmichael2009} 
(simplifications while applying this procedure can be achieved by using the 
properties of $\mathbb{U}_{}^{}(t)$) 
which when Fourier transformed back, we get 
$\bar{\mathbb{P}}_{}^{}[\bm{\Upsilon}_{}^{},t]$. Using thus obtained solution, 
$\mathbb{P}_{}^{}[\bm{\Upsilon}_{}^{},t]$ is given as,
\begin{eqnarray}
\label{eq-s7}
\mathbb{P}_{}^{}[\bm{\Upsilon}_{}^{},t]&=&\int_{}^{}\mathcal{D}[\bm{\Upsilon}_{}
^{\prime}]\
\mathbb{G}_{}^{}[\bm{\Upsilon}_{}^{},t_{}^{}|\bm{\Upsilon}_{}^{\prime},0]\mathbb
{P}_{}^{}[\bm{\Upsilon}_{}^{\prime},0]
\end{eqnarray}
with the Greens function or the propagator given by,
 \begin{eqnarray}
  \label{eq-s8}
  &&\mathbb{G}_{}^{}[\bm{\Upsilon}_{}^{},t_{}^{}|\bm{\Upsilon}_{}^{\prime},0] = 
\frac{1}{\pi_{}^{M}}\frac{e^{\frac{1}{2} \mathbf{Tr}[\Gamma_{}^{}] 
t_{}^{}}_{}}{\sqrt{\mathbf{Det}\left[\mathbb{U}_{21}^{}(t)\bm{\sigma}_{x}^{}
\right]}} \times \nonumber\\
  &&e^{-\frac{1}{2} \left\{ \bm{\Upsilon}_{}^{T}  
\left[\mathbb{U}_{12}^{}(t){\mathbb{U}_{22}^{}(t)}_{}^{-1}\right] 
\bm{\Upsilon}_{}^{}+\left[\bm{\Upsilon}_{}^{} -\mathbb{U}_{22}^{}(t) 
\bm{\Upsilon}_{}^{\prime} \right]_{}^{T} 
\left[\mathbb{U}_{21}^{}(t){\mathbb{U}_{22}^{}(t)}_{}^{T}\right]_{}^{-1} 
  \left[\bm{\Upsilon}_{}^{} -\mathbb{U}_{22}^{}(t) \bm{\Upsilon}_{}^{\prime} 
\right]_{}^{}\right\}}_{}\nonumber\\
  \end{eqnarray}
and $\int_{}\mathcal{D}[\bm{\Upsilon}_{}^{\prime}]=$\\ 
$\int_{-\infty}^{+\infty}d[\Re(\gamma_{1}^{\prime})\int_{-\infty}^{+\infty}d[
\Im(\gamma_{1}^{\prime})]\cdots\int_{-\infty}^{+\infty}d[\Re(\gamma_{M}^{\prime}
)\int_{-\infty}^{+\infty}d[\Im(\gamma_{M}^{\prime})]$.
 \subsubsection{Method II}
 The formal solution of the parabolic partial differential 
equation \cite{Dattoli1997},
  \begin{eqnarray}
   \label{eq-s9}
   \frac{\partial}{\partial 
t}\mathbb{P}_{}^{}[\bm{\Upsilon}_{}^{},t]&=&\frac{1}{2}
  \left[ \begin{pmatrix} \bm{\Upsilon}_{}^{} \\ 
\bm{\nabla}_{\bm{\Upsilon}_{}^{}}^{} \end{pmatrix}^{T}_{}
  \mathbb{H}_{}^{}
  \begin{pmatrix} \bm{\Upsilon}_{}^{} \\ \bm{\nabla}_{\bm{\Upsilon}_{}^{}}^{} 
\end{pmatrix}^{}_{}
  + \mathbf{Tr}[\Gamma_{}^{}]\right] 
\mathbb{P}_{}^{}[\bm{\Upsilon}_{}^{},t]\nonumber\\
 \end{eqnarray}
 is
 \begin{eqnarray}
  \label{eq-s10}
  &&\mathbb{P}_{}^{}[\bm{\Upsilon}_{}^{},t]=e_{}^{\frac{1}{2}
  \left[ \begin{pmatrix} \bm{\Upsilon}_{}^{} \\ 
\bm{\nabla}_{\bm{\Upsilon}_{}^{}}^{} \end{pmatrix}^{T}_{}
  \mathbb{H}_{}^{}
  \begin{pmatrix} \bm{\Upsilon}_{}^{} \\ \bm{\nabla}_{\bm{\Upsilon}_{}^{}}^{} 
\end{pmatrix}^{}_{}
  + \mathbf{Tr}[\Gamma_{}^{}]\right] t}\mathbb{P}_{}^{}[\bm{\Upsilon}_{}^{},0].
 \end{eqnarray}
 The exponential operator in the above equation can be put in a more manageable 
form using Wei-Norman method \cite{Wei1963,Wei1964} inspired technique 
\cite{Wang1998} as,
 \begin{eqnarray}
  \label{eq-s11}
  &&e_{}^{\frac{1}{2}
  \left[ \begin{pmatrix} \bm{\Upsilon}_{}^{} \\ 
\bm{\nabla}_{\bm{\Upsilon}_{}^{}}^{} \end{pmatrix}^{T}_{}
  \mathbb{H}_{}^{}
  \begin{pmatrix} \bm{\Upsilon}_{}^{} \\ \bm{\nabla}_{\bm{\Upsilon}_{}^{}}^{} 
\end{pmatrix}^{}_{}\right] t} = \frac{1}{\sqrt{\mathbb{U}_{22}^{}(t)}} \times 
\nonumber\\  
&&e_{}^{-\frac{1}{2}\bm{\Upsilon}_{}^{T}\left[\mathbb{U}_{12}^{}(t){\mathbb{U}_{
22}^{}(t)}_{}^{-1}\right]\bm{\Upsilon}_{}^{}}
  e_{}^{-\bm{\Upsilon}_{}^{T} 
\left[\ln{\mathbb{U}_{22}^{}(t)}_{}^{T}\right]\bm{\nabla}_{\bm{\Upsilon}_{}^{}}^
{}}
  e_{}^{\frac{1}{2} \bm{\nabla}_{\bm{\Upsilon}_{}^{}}^{T} 
\left[{\mathbb{U}_{22}^{}(t)}_{}^{-1}\mathbb{U}_{21}^{}(t)\right] 
\bm{\nabla}_{\bm{\Upsilon}_{}^{}}^{}}\nonumber\\
 \end{eqnarray}

 where $\mathbb{U}_{xy}^{}(t)$ are same as defined previously. Using this, 
$\mathbb{P}_{}^{}[\bm{\Upsilon}_{}^{},t]$ can be expressed  \cite{Dattoli1997} 
in the same form as given previously. 
\subsection{Counting field independent Wigner function of the system}
The Wigner function of the system for the case 
$\bm{\chi}_{}^{}=\bm{\lambda}_{}^{}=\mathbf{0}_{}^{}$, i.e., in the absence of 
two-point measurements, is given as,
\begin{eqnarray}
\label{eq-s12} 
\mathbb{P}_{}^{}[\bm{\Upsilon}_{}^{},t]&=&\frac{1}{\pi_{}^{M}}\frac{1}{\sqrt{
\textbf{Det}\left[\mathbf{D}_{S}^{}(t)\bm{\sigma}_{x}^{}\right]}}e_{}^{-\frac{1}
{2} \bm{\Upsilon}_{}^{T}\mathbf{D}_{S}^{}(t)\bm{\Upsilon}_{}^{}},\nonumber\\
\end{eqnarray}
with
\begin{eqnarray}
 \label{eq-s13}
 \mathbf{D}_{S}^{}(t) &=& e_{}^{-\frac{1}{2} \left[\displaystyle{\sum_{\alpha = 
1}^{N}\mathbf{\Gamma}_{\alpha}^{}}\right] t}\mathbf{D}_{S}^{} e_{}^{-\frac{1}{2} 
\left[\displaystyle{\sum_{\alpha = 1}^{N}\mathbf{\Gamma}_{\alpha}^{}}\right] 
t}\nonumber\\
 &&+ \displaystyle{\int_{0}^{t} ds}e_{}^{-\frac{1}{2} 
\left[\displaystyle{\sum_{\alpha = 1}^{N}\mathbf{\Gamma}_{\alpha}^{}}\right] s} 
\left[\sum_{\alpha = 
1}^{N}\mathbf{\Gamma}_{\alpha}^{}\mathbf{D}_{\alpha}^{}\right] 
e_{}^{-\frac{1}{2} \left[\displaystyle{\sum_{\alpha = 
1}^{N}\mathbf{\Gamma}_{\alpha}^{}}\right] s}.\nonumber\\
\end{eqnarray}
\subsection{\protect{Expressions for the matrix elements of $\mathbb{X}$}}
The expressions for the elements of the matrix, 
$\mathbb{X}_{}^{}[\bm{\chi}_{}^{},\bm{\lambda}_{}^{}] = 
\begin{pmatrix}\mathbb{X}_{--}^{}[\bm{\chi}_{}^{},\bm{\lambda}_{}^{}] & 
\mathbb{X}_{-+}^{}[\bm{\chi}_{}^{},\bm{\lambda}_{}^{}] \\ 
\mathbb{X}_{+-}^{}[\bm{\chi}_{}^{},\bm{\lambda}_{}^{}] & 
\mathbb{X}_{++}^{}[\bm{\chi}_{}^{},\bm{\lambda}_{}^{}]\end{pmatrix}$, are given 
as,
\begin{widetext}
\begin{eqnarray}
 \label{eq-s16}
 &&\mathbb{X}_{--}^{}[\bm{\chi}_{}^{},\bm{\lambda}_{}^{}] = \frac{1}{2} - 
\frac{1}{2}\frac{1}{\left({\Lambda_{-}^{}[\bm{\chi}_{}^{},\bm{\lambda}_{}^{}]}_{
}^{2}-{\Lambda_{+}^{}[\bm{\chi}_{}^{},\bm{\lambda}_{}^{}]}_{}^{2}\right)_{}^{2}}
 \textbf{Det}\left[\displaystyle{\underset{\alpha \neq 
\alpha^\prime}{\sum_{\alpha, \alpha^\prime = 1, 
2}^{}}\left\{\left(\frac{{\Lambda_{-}^{}[\bm{\chi}_{}^{},\bm{\lambda}_{}^{}]}_{}
^{2}+{\Lambda_{+}^{}[\bm{\chi}_{}^{},\bm{\lambda}_{}^{}]}_{}^{2}}{2}\right) 
I_{}^{} - \Xi_{\alpha_{}^{} 
\alpha_{}^{\prime}}[\bm{\chi}_{}^{},\bm{\lambda}_{}^{}]\right\}} + 
\Xi_{12S}^{}[\bm{\chi}_{}^{},\bm{\lambda}_{}^{}]\right],\nonumber\\
 &&\mathbb{X}_{\mp\pm}^{}[\bm{\chi}_{}^{},\bm{\lambda}_{}^{}] =\nonumber\\
 &&\frac{1}{2} \textbf{Tr}\left[\displaystyle{\sum_{\alpha = 1, 
2}^{}\Xi_{\alpha 
S}^{}[\chi_{\alpha}^{},\lambda_{\alpha}^{}]}\right] \pm 
\frac{1}{\left({\Lambda_{-}^{}[\bm{\chi}_{}^{},\bm{\lambda}_{}^{}]}_{}^{2}-{
\Lambda_{+}^{}[\bm{\chi}_{}^{},\bm{\lambda}_{}^{}]}_{}^{2}\right)_{}^{}}
 \textbf{Tr}\left[\displaystyle{\underset{\alpha \neq 
\alpha^\prime}{\sum_{\alpha, \alpha^\prime = 1, 
2}^{}}\left\{\left(\frac{{\Lambda_{-}^{}[\bm{\chi}_{}^{},\bm{\lambda}_{}^{}]}_{}
^{2}+{\Lambda_{+}^{}[\bm{\chi}_{}^{},\bm{\lambda}_{}^{}]}_{}^{2}}{2}\right) 
I_{}^{} - \Xi_{\alpha_{}^{} 
\alpha_{}^{\prime}}[\bm{\chi}_{}^{},\bm{\lambda}_{}^{}]\right\}\Xi_{\alpha 
S}^{}[\chi_{\alpha}^{},\lambda_{\alpha}^{}]}\right]\nonumber\\
 &&\text{and}\nonumber\\
 &&\mathbb{X}_{++}^{}[\bm{\chi}_{}^{},\bm{\lambda}_{}^{}] = 
\textbf{Det}\left[\displaystyle{\sum_{\alpha = 1, 2}^{}\Xi_{\alpha 
S}^{}}[\chi_{\alpha}^{},\lambda_{\alpha}^{}]\right] + 
\left(\frac{{\Lambda_{-}^{}[\bm{\chi}_{}^{},\bm{\lambda}_{}^{}]}_{}^{2}+{
\Lambda_{+}^{}[\bm{\chi}_{}^{},\bm{\lambda}_{}^{}]}_{}^{2}}{2}\right) 
\left(1-\mathbb{X}_{--}^{}[\bm{\chi}_{}^{},\bm{\lambda}_{}^{}]\right)
\end{eqnarray}
 with
\begin{eqnarray}
 \label{eq-s17}
 \Xi_{12S}^{}[\bm{\chi}_{}^{},\bm{\lambda}_{}^{}] &=& \Gamma_{1}^{} 
\Gamma_{2}^{} \left[e_{}^{i \epsilon_{}^{} \lambda_{1}^{} 
\sigma_{z}^{}}\sigma_{x}^{}D_{1}^{}e_{}^{-i \epsilon_{}^{} \lambda_{1}^{} 
\sigma_{z}^{}},e_{}^{i \epsilon_{}^{} \lambda_{2}^{} 
\sigma_{z}^{}}\sigma_{x}^{}D_{2}^{}e_{}^{-i \epsilon_{}^{} \lambda_{2}^{} 
\sigma_{z}^{}}\right]\times\nonumber\\
 &&\left\{\left[\sigma_{x}^{}D_{S}^{}-\frac{1}{2} 
I_{}^{}\right]\left[\left(e_{}^{-i\epsilon_{}^{}\chi_{1}^{}}-1\right)-\left(e_{}
^{-i\epsilon_{}^{}\chi_{2}^{}}-1\right)\right]\left[\left(e_{}^{i\epsilon_{}^{}
\chi_{1}^{}}-1\right)+\left(e_{}^{i\epsilon_{}^{}\chi_{2}^{}}-1\right)\right]
\right.\nonumber\\
 &&\left.- \left[\sigma_{x}^{} D_{S}^{}+\frac{1}{2} 
I_{}^{}\right]\left[\left(e_{}^{i\epsilon_{}^{}\chi_{1}^{}}-1\right)-\left(e_{}^
{i\epsilon_{}^{}\chi_{2}^{}}-1\right)\right]\left[\left(e_{}^{-i\epsilon_{}^{}
\chi_{1}^{}}-1\right)+\left(e_{}^{-i\epsilon_{}^{}\chi_{2}^{}}-1\right)\right]
\right\}\nonumber\\
\end{eqnarray}
\end{widetext}
and $\Xi_{\alpha S}^{}[\chi_{\alpha}^{},\lambda_{\alpha}^{}]$ is given in Eq. 
(\ref{eq-24}).
\section*{References}
\bibliography{statistics_of_energy_transport_across_squeezed_reservoirs.bib}
\end{document}